\documentclass[11pt,letterpaper]{article}
\usepackage[latin1]{inputenc}
\usepackage{float}
\usepackage{textcomp}
\usepackage{amsmath}
\usepackage{amssymb}
\usepackage{graphicx}
\usepackage[authoryear]{natbib}

\makeatletter

\providecommand{\tabularnewline}{\\}

\usepackage{enumerate}
\usepackage{amsfonts}

\makeatother

\begin{document}

\title{On the Efficient Market Hypothesis of Stock Market Indexes: \\
The Role of Non-synchronous Trading and Portfolio Effects}

\author{R. Ortiz%
\thanks{Faculty of Engineering, Universidad Diego Portales, Chile. Email:
roberto.ortiz@udp.cl%
}, M. Contreras%
\thanks{Faculty of Engineering \& Sciences, Universidad Adolfo Ibáñez, Chile.
Email: mauricio.contreras@uai.cl%
} and M. Villena%
\thanks{Faculty of Engineering \& Sciences, Universidad Adolfo Ibáñez, Chile.
Email: marcelo.villena@uai.cl%
}}
\maketitle
\begin{abstract}
In this article, the long-term behavior of the stock market index
of the New York Stock Exchange is studied, for the period 1950 to
2013. Specifically, the CRSP Value-Weighted and CRSP Equal-Weighted
index are analyzed in terms of market efficiency, using the standard
ratio variance test, considering over 1600 one week rolling windows.
For the equally weighted index, the null hypothesis of random walk
is rejected in the whole period, while for the weighted market value
index, the null hypothesis start to be accepted from the 1990s. In
order to explain this difference, we raised the hypothesis that this
behavior can be explained by the joint action of portfolios and non-synchronous
trading effects. To check the feasibility of the above assumption,
we performed a simulation of both effects, on two- and six-asset portfolios.
The results showed that it is possible to explain the empirical difference
between the two index, almost entirely by the joint effects of portfolio
and non-synchronous trading. 
\end{abstract}
\vspace{1cm}

\pagebreak{}

\section{Introduction}

The efficient financial markets hypothesis developed by \citet*{fama1970efficient}
established that no investor could profit in excess in a systematic
way using public or private information. That is, asset prices instantly
reflect all information about financial assets in the capital market.
In order to prove this hypothesis, transactions and information costs
must be zero at all times {[}\citet{grossman1980impossibility}{]}.
Since, there will always be transaction and information costs, different
versions of the efficient market hypothesis have been introduced:
the weak, semi-strong, and strong ways, see \citet*{Robert1967}%
\footnote{The weak way implies that the asset prices reflect all past information
obtained from the historical series of prices. The semi-strong way
assumes that prices reflect the information contained in the historical
series and also all the public information related to the price determination
of financial assets. And finally, the strong way implies that prices
reflect all the information kept in the historical series of prices,
all public information, and all private information about the financial
asset in the analysis.%
}. Provided that transactions and information costs are definitely
higher than zero, the strong version of efficient capital market hypothesis
is always false {[}\citet{fama1991efficient}{]}. For this reason,
there is a growing agreement that, for all practical purposes, only
the weak version of an efficient market is feasible (see for example
\citet{fama1965behavior}, \citet{jensen1978some}, \citet{lo1988stock},
\citet{fama1991efficient}, \citet{campbell1997econometrics}). This
weak version states that no investor can profit on excess by taking
a position in the market only observing the behavior of past prices
of financial assets. \\

It is often argued that, if there are some investors with more information
than that contained in the historical data and with a higher ability
to analyze it, then these investors will be capable of getting a higher
return than others who only take into account historical info. However,
the additional benefit obtained can involve more costs due to the
operation of obtaining and processing this extra data. In this way,
the net return earned is virtually null. Thus, this weak version has
a higher economic sense, because it states that prices reflect (at
every moment) the point where the marginal benefit of using the information
does not exceed the marginal cost of obtaining this same information
{[}\citet{jensen1978some}{]}.\\

Nevertheless, the existence of transactions and information costs
is not the main obstacle to making inferences about market efficiency.
This hypothesis must be necessarily be tested with a price equilibrium
model, that is, a model for the evaluation of financial assets. Therefore,
if there is evidence of an anomalous behavior of returns, we must
be wary of whether the culprit is the market failure or an inaccuracy
in the proposed equilibrium price model {[}\citet{fama1970efficient}{]}.
\\

Within the tests that allow evaluation of the efficiency market hypothesis,
the ratio variance test proposed by \citet{lo1988stock} is highlighted.
This test is based on the following argument: If a series of logarithms
of prices shows no linear autocorrelation, then the variance of the
sum of k periods must be equal to k times the variance of one period.
Thus, the random walk hypothesis can be measured by comparing data
variances from different time spans.\\

Applications of the ratio variance test in the decade of the 1980s
had a strong influence on the acceptance of the claim that financial
asset prices do not follow a random walk movement. This test was designed
to be robust to a variety of types of heteroskedasticity and not normality.
In fact, \citet{lo1988stock} in a very influential work reported
that the null hypothesis of market efficiency in the form of random
walk was rejected for the CRSP NYSE Value-Weighted and CRSP NYSE Equal-Weighted
index for a sample of weekly data from September 6, 1962, to December
26, 1985.\\

Nearly 25 years after Lo and MacKinlay's work, we looked into these
conclusions by applying the variance ratio test to the main stock
market indexes, but this time by considering a larger weekly return
data sample, from 1950 to 2013. In order to appreciate the evolution
of results for this test we used 1600 one week rolling windows. The
results obtained using this entire period indicated that the random
walk hypothesis could be ruled out for the CRSP NYSE Equally Weighted
index. In contrast, for the CRSP NYSE Value Weighted index, the random
walk hypothesis started to not be rejected, with growing strength
from the 1990s. \\

Note that, given the CRSP NYSE Value Weighted Index tended to be more
representative of the way a diversified person invests, the above
evidence has significant future implications for investors. Since
these results are different from conclusions given by Lo and MacKinlay
(1988), we carried out a study to show the evolution of behavior of
these indexes in time. In this way, to be sure that these conclusions
are reliable, we applied a test to estimate the empirical distribution
of the statistics by using the bootstrapping method {[}\citet{kim2006wild}{]}.
This last test overcame the drawbacks of finite samples and estimates
of variance with overlapping. The conclusions obtained using this
last test were the same.\\

Naturally, the following question emerges: Why do two indexes built
on the same assets show different responses to the random walk test?
In order to explain the above effects, we carry out an analysis of
the combined effects of non-synchronous trading and portfolio asset
location over the variance ratio test of the random walk hypothesis.
The results show that the acceptance/rejection of the random walk
hypothesis depends on:
\begin{enumerate}
\item The asset structure in which the portfolio is built and 
\item The non-trading possibilities of the same goods. 
\end{enumerate}
Even when the individual asset returns have no autocorrelations, the
combined effects of non-synchronous trading and portfolio can induce,
in some cases, positive autocorrelations when the correlation between
various assets is positive. These spurious autocorrelations thus can
explain the different behaviors for the ratio variance test when it
is applied to equal-weighted and value-weighted index.\\

Here, the modeling and simulation of both the non-trading effect and
the portfolio effect is executed and the results show that the combined
effects can cause a rejection of the null hypothesis for some portfolios.
The positive autocorrelation in some index is produced by the update
of the return of assets with non-trading problems in relation to the
return of normal trading assets. \\

The article is organized in the following way: in section 2, a review
of literature is presented. Section 3, it gives the specifications
of the statistical model with which the ratio-variances test is performed.
Section 4 analyzes the evolution of the ratio variance test over the
principal index of the New York Exchanges. In section 5, we explore
the possibility that the non-synchronous trading and portfolio effects
explain the divergence between the results of equally weighted index
and index weighted by market value. Finally, the conclusions are delivered
in section 6.

\section{Literature Review}

In the traditional capital assets pricing theory, it is proposed that
for a higher non diversifiable risk, a higher return should be obtained.
It is assumed that investors will receive on average a return proportional
to the non diversifiable risk that they take. In practice, the real
returns fluctuate randomly around their expected value. Differences
between returns realized and expected are unpredictable, and in this
sense they follow random walks. If this is true, then the capital
market hypothesis in its weak version is not refused {[}\citet{campbell1997econometrics}{]}.\\

The hypothesis that the returns of assets follow a random walk suggests
that no one can systematically predict changes in the asset prices
in order to obtain higher returns. Whether the changes in prices are
predictable based on past price changes (or whether changes in prices
do not follow a random walk) is still a subject of controversy and
empirical research.\\

In the literature on the random walk theory, three types of hypothesis
have been suggested {[}\citet{campbell1997econometrics}{]}: Random
Walk 1 (RW1), Random Walk 2 (RW2), and Random Walk 3 (RW3). \\

\begin{description}
\item [{i)}] The RW1 hypothesis assumes that continuously composed returns
are independent and identically distributed. 
\item [{ii)}] The RW2 hypothesis assumes that returns are independent and
allow that the distribution of assets changes over time. 
\item [{iii)}] The RW3 hypothesis is the most general and implies that
returns in different periods are not linearly correlated. 
\end{description}
It is typical in the latter case to assume processes do not have linear
dependence but have quadratic dependency asymptotically decreasing
to zero. For example, stochastic models with heteroskedasticity, such
as GARCH processes, can be considered {[}\citet{engle1982autoregressive}
and \citet{bollerslev1986generalized}{]}.

\subsection{On the ratio variance test}

The ratio variance test used to evaluate the RW3 hypothesis is built
on the following argument: If a series of returns compounded continuously
do not show linear autocorrelation, then the variance of the sum of
returns of $k$ periods must be equal to $k$ times the variance of
one period return. \\
 For example, consider the accumulated returns of two periods: 
\begin{equation}
r_{2t}=r_{t}+r_{t-1}
\end{equation}
Then the variance ratio $VR(2)$ is defined by 
\begin{equation}
VR(2)=\frac{V(r_{2t})}{2V(r_{t})}=\frac{V(r_{t})+V(r_{t-1})+2Cov(r_{t},r_{t-1})}{2V(r_{t})}
\end{equation}
can be used as a measure of the efficiency with which the time series
satisfies the random walk scenario. In addition, if the process is
stationary, that is, $V(r_{t})=V(r_{t-1})$ and $\rho_{1}=\frac{Cov(r_{t},r_{t-1})}{V(r_{t})}$,
then 
\[
VR(2)=1+\rho_{1}
\]
Note that if the time series satisfies RW3, then VR(2) is equal to
1. \\
 In general, for $k$ periods it is obtained that 
\[
VR(k)=1+2\sum_{i=1}^{k-1}(1+\frac{i}{k})\rho_{i}
\]
If the autocorrelations are higher than zero, the variance ratio is
greater than one, and if the autocorrelations are negative, the variance
ratio is less than one. The variance ratio test then seeks to determine
where VR(k) is different from one.\\

In the work of \citet{lo1988stock}, the variance ratio test became
a relevant statistical test. They proved the RW3 hypothesis using
a sample of weekly returns covering the period between 1962 and 1985.
They applied this test to two of the main index of the New York stock
market and a set of different portfolios sorted by size. They concluded
the random walk hypothesis was rejected for each of the considered
series.\\

In order to perform statistical inference from the statistic VR(k),
certain assumptions about the return generating process must be made,
and these assumptions must be consistent with the observed empirical
distributions of returns. In general, the assets return time series
showed time-dependent volatility and that the empirical distributions
of returns show leptokurtosis.\\

The above-mentioned aspects are in the proposal of the variance ratio
test considered by \citet{lo1988stock}. In fact, they explicitly
assume that returns are generated by a process of $\phi$ mixing {[}\citet{white1980heteroskedasticity}{]}.
The $\phi$ mixing assumption allows making inferences by using the
asymptotic distribution of the statistic. \\

It should be noted here that the inference proves all these assumptions
in a joint and unfolding way. Therefore, if the random walk hypothesis
is rejected, it might be because the process followed by return is
not a random walk one or because some of the model hypothesis assumed
for generating the returns process is wrong.\\

The VR test proposed by Lo and MacKinlay (1988) has other problems,
such as:

a) The VR test is an asymptotic test; that is, it is valid in principle
for an infinite sampling data size. Hence, it cannot be reliable for
small finite sampling.

b) The VR test typically uses overlapping data to compute the variance
of the long-horizon returns. It adds difficulty to analyzing the exact
statistical distribution of the sample variance ratio.\\
 \\
 After the contribution by \citet{lo1988stock}, certain improvements
of the inference of the VR test were proposed to surpass the above-mentioned
issues. In fact, a number of distinct tests done with samples of finite
sizes were proposed, for example:
\begin{itemize}
\item A modified ratio variance test {[}\citet{chen2006variance}{]}. 
\item A test based on ranking and sign changing {[}\citet{wright2000alternative}{]}. 
\item A test that uses subsample methods {[}\citet{whang2003multiple}{]}. 
\item The bootstrapping method {[}\citet{kim2006wild}{]}. 
\end{itemize}
These tests have been used to estimate the empirical distribution
of ratio-variance statistics and to improve the inference process.
On the other hand, the test by \citet{lo1988stock} is considered
a simple test that can be applied in time intervals of different spans
(for example 2, 4, 8, and 16 week periods). Observe, that it is enough
the test is rejected for one of these intervals, in order that the
random walk hypothesis is to be turned away. Due that these tests
are applied independently and at the same time, this procedure leads
to an oversized test. Therefore, multiple comparison tests have been
proposed {[}\citet{chow1993simple}{]}, {[}\citet{richardson1991tests}{]},
{[}\citet{cecchetti1994variance}{]}.\\

\subsection{On the nature of distribution of returns }

Another issue in which literature has paid attention is, regarding
the random walk hypothesis, it is the nature of the empirical probability
distribution of price changes in financial assets. It is an essential
point, given that the nature of return distributions affects both,
the statistical test types used in the research and the interpretation
of the obtained results {[}\citet{fama1970efficient}{]}, {[}\citet{kan2006exact}{]},
{[}\citet{kim2006wild}{]}.\\

One of the first models to study the evolution followed by financial
assets prices was developed by \citet{bachelier1900theorie}. This
model assumed that returns of financial assets are identically and
independently distributed with a normal distribution. Here, the price
changes are caused by the sum of a vast number of independent variables
associated with the information used by investors. They can be thought
of as a set of decisions in statistical equilibrium, with properties
relatively similar to a set of particles in statistical mechanics
{[}\citet{osborne1959brownian}{]}. One implication of this is that,
when applying the central limit theorem, the compound returns are
normally distributed. However, the empirical distribution of the price
logarithms does not support the hypothesis of normality.\\

For the empirical distribution, it is typical to observe queues greater
than those predicted for a normal distribution. That is, the kurtosis
is significantly higher than 3, and the presence of bias is also observed.
It has been suggested that this behavior could be explained by a more
general family of distributions. Specifically, it is proposed that
changes in the logarithms of prices of financial assets can be represented
by Pareto stable distributions {[}\citet{mandelbrot1997variation}{]}.\\

This type of distribution includes the normal distribution as a special
case and allows consideration of both the leptokurtosis and the bias
observed in the empirical distributions. Different empirical investigations
have concluded that Pareto stable distributions represent a better
description of the daily returns than the normal distribution {[}\citet{fama1965behavior}{]},
{[}\citet{kanellopoulou2008empirical}{]}.\\

The family of Pareto stable distributions is characterized by four
parameters: The $\alpha$ parameter that determines the shape of the
distribution, the $\beta$ parameter associated with the bias, the
$\gamma$ parameter related to the scale, and the $\delta$ parameter
linked to the location. When $\alpha=2$ and $\beta=0$, the distribution
is normal. The Maldelbrot hypothesis (1963) states that empirical
returns of financial assets can be described by Pareto stable distributions,
with the $\alpha$ parameter taking values between 1 and 2. The Pareto
stable distributions have two important properties: \\

\begin{description}
\item [{a)}] They have stability or invariability under the addition; and 
\item [{b)}] These distributions are the only asymptotic distributions
for the sum of independent and identically distributed random variables
{[}\citet{fama1965behavior}{]}. 
\end{description}
However, a major problem arises if the compound returns are distributed
as Pareto stable distributions. In this case, in general, the estimated
parameter $\alpha$ is less than 2, and the distribution has an infinite
variance. Therefore, many of the commonly used statistical tools that
assume finite variance, would not provide any conclusions.\\

In spite of this, most researchers have assumed that continuously
compound returns of financial assets have short-term dependencies
and finite variance, contrary to the behavior of the Pareto stable
distribution processes (for example see, \citet{atchison1987nonsynchronous},
\citet{boudoukh1994tale}, \citet{lo1988stock} and \citet{white1984nonlinear}).\\

It is common to assume that asset returns can be represented as arising
from overlapped mixtures of normal distributions. In this case, returns
may have normal conditional distributions; some concentrated around
the average with lower variance and others with greater variance that
put more weight on the distribution queues.\\

These mixed distributions can explain the observed empirical unconditional
distributions that show greater queues than those predicted by a normal
distribution. Given that every moment of these distributions is finite,
the central limit theorem applies, and asymptotically the long-term
distribution will be normal {[}\citet{campbell1997econometrics}{]}.
This last argument is very important in making inferences regarding
the parameters of each model, because it allows determination of the
distribution of asymptotic probabilities of each parameter.

\subsection{On non-synchronous trading }

It is an empirical fact that the null hypothesis is rejected with
more force by portfolios whose small firms stock are more heavily
weighted than the greater firms stock (see for example \citet{atchison1987nonsynchronous},
\citet{perry1985portfolio}, and \citet{cohen1983friction}).\\

Small firms are characterized by the fact that they have trading that
is not synchronized, so different behavior of the CRSP Value-Weighted
and CRSP Equal-Weighted from the 1990s could rest on this fact. This
has led researchers to consider the effect of a non-synchronous trading
over the output of the variance ratio test.\\

The non-synchronous trading is an effect associated with two or more
time series. As a result, it appears that the transactions or movements
in the prices of different assets are updated at different time intervals.\\

Consider an asset time series that is recorded at fixed time intervals.
In general, the price time series is updated each time interval as
a result of a subjacent trading process. But typically for small companies,
this process is not necessarily done at each interval. In this way,
the data is not frequently updated by the trading process. This last
situation is called a non-trading effect. For example, in a daily
price series, normally the registered daily price is the closing price.
This closing price is normally the toll of the last transaction made
during the day. Also, if there is no asset trading during that day,
the end price is then the closing price of the previous day.\\

The non-trading price of an individual asset and the fact that some
assets register their price changes non-synchronously can modify the
estimation of statistical parameters. For instance, the individual
asset autocorrelations or cross-autocorrelations between individual
assets are modified, among other variables.\\

Non-synchronous trading may induce a significant spurious correlation
in stock returns and create a false impression of predictability in
asset returns. It also affects the results of the variance ratio test
(see for example, \citet{mech1993portfolio}, \citet{perry1985portfolio},
\citet{campbell1997econometrics} and \citet{lo1988stock}).\\

Theoretical models have been proposed to estimate the autocorrelation
based only on the non-synchronous trading effect. These findings indicate
that, the theoretical impact is significantly lower than that which
has been empirically observed. In this way, this effect alone cannot
explain the rejection of the random walk null hypothesis. (see for
example \citet{atchison1987nonsynchronous}, \citet{conrad1988time},
\citet{lo1990econometric} and \citet{mech1993portfolio}). These
models suppose that the individual assets have homogeneous characteristics,
that is, the non-trading probabilities associated with different assets
are identical.\\

However, in other research, heterogeneous individual asset characteristics
have been considered. For instance, \citet{boudoukh1994tale} used
the model of \citet{scholes1977estimating} to estimate the theoretical
autocorrelation induced by non-synchronous trading. Here, the non-trading
probabilities were different for each asset included in the portfolio.
They estimated that the autocorrelation induced by non-synchronous
trading for equally weighted portfolios could reach a 17.82\%.\\

These models have also been used to value the impact of non-synchronous
trading over the statistical inference process. Specifically, the
focus in \citet{lo1988stock} and \citet{lo1990econometric} was to
determine whether non-synchronous trading can lead to rejection of
the random walk null hypothesis. These authors concluded that this
effect is so negligible to be the cause of a null hypothesis rejection.
It is important to note that these authors considered:
\begin{itemize}
\item Asymptotic estimation of the parameters 
\item Infinite portfolio asset size 
\item A particular price model as valid and 
\item In general, individual homogeneous features assets. 
\end{itemize}
All of the above-mentioned studies considered an asymptotic behavior
and assumed that the portfolios were efficient. Therefore, there was
only systematic risk. However, in practice, statistical inference
should be performed with finite-sized samples. In this case, the values
of the parameters and their associated errors were estimated with
the assumption that the pricing model used was the correct one. This
simplified the analysis, but made the inference outcomes dependent
on the accuracy of the price model.

\subsection{The portfolio effect}

The properties of an index depends on the multivariate process followed
by financial assets that constitute them and the individual weight
assigned to each asset.\\

Consider the typical case in which it is assumed that the return of
assets follow a multivariate normal process $N(\mu,V)$, where $\mu$
is the vector of expected returns of assets and $V$ is the covariance
matrix between return of assets. Therefore, to study the behavior
of an asset portfolio, we must estimate $n$ values of the average
returns and the $\frac{n(n-1)}{2}$ values of the covariance matrix
$V$. \\

When there is non-synchronous trading, the potential bias in the estimation
of parameters are added to the estimation problem. Risk estimation
and non-synchronous trading could have a strong impact on the performance
of asset allocation models.\\

In addition, when there is non-synchronous trading, crossed autocorrelations
between assets which are detectable even though the subjacent correlation
between assets is just contemporary {[}\citet{fisher1966some}{]}.
This effect is intensified when the observation interval decreases.
On the other hand, it has been reported that when the observation
period decreases, approaching zero, correlations between assets tends
to be zero {[}\citet{epps1979comovements}{]}.\\

Diverse studies have explored the effects that non-synchronous trading
may have on portfolio autocorrelation. \citet{scholes1977estimating},
\citet{dimson1979risk}, and \citet{cohen1983friction} showed that
the non-synchronous trading effect causes an underestimation of the
betas in financial assets. They also proposed methods to mitigate
this effect. Later, \citet{atchison1987nonsynchronous} and \citet{lo1990econometric}
estimated the theoretical correlation induced in portfolios by non-synchronous
trading and concluded that it was significantly lower than which had
been empirically observed. In particular, Atchinson et al. (1987)
estimated the theoretical non-trading induced autocorrelation to be:
\\

a) 4\% for the NYSE Equal-Weighted Index, and

b) 2\% for the NYSE Value-Weighted Index.\\
 \\
 The observed empirical correlation values were 28\% for the NYSE
Equal Weighted index and 16\% for the NYSE Value Weighted index. Similar
results were reported by \citet{lo1990econometric}. \\

For the estimation of portfolio autocorrelations, in the above models
it was assumed that the portfolio contained infinite assets and that
it was entirely diversified. Recently, \citet{chelley2014portfolio}
conducted research on the effects of non-trading on autocorrelations
considering portfolios of finite size. They concluded that the autocorrelation
induced by non-trading was influenced by non-systematic risks.\\

\section{The Model: Specification of the Ratio Variance Test}

In the section following, we review the ratio-variance test developed
by \citet{lo1988stock}. \\

Let $P_{t}$ be the price of a financial asset and $X_{t}=\ln(P_{t})$
the process followed by its logarithm at time t. The null hypothesis
that $X_{t}$ follows a random walk can be expressed through the following
recursive relationship: 
\begin{eqnarray}
X_{t}=\mu+X_{t-1}+\varepsilon_{t}
\end{eqnarray}
where $\mu$ is a parameter of arbitrary tendency and $\varepsilon_{t}$
is a random perturbation. It is assumed that for every value of t
the error expected value is zero {[}$E(\varepsilon_{t})=0${]}.\\

An important property of a random walk $X_{t}$ is that its variance
grows linearly with the observational time length. That is, the variance
of the return of the two periods is twice the variance of one period.
This feature is essential to prove the hypothesis of the random walk.
It is assumed that time intervals are equally spaced and that there
are (2n + 1) observations $[X_{0},X_{1},.,X_{2n}]$ of $X_{t}$ in
each interval. Therefore, the average $\mu$ and variance $\sigma^{2}$
can be estimated through the relations:

\begin{equation}
\hat{\mu}=\frac{1}{2n}\sum_{k=1}^{2n}(X_{k}-X_{k-1})=\frac{1}{2n}(X_{2n}-X_{0})
\end{equation}

\begin{equation}
\hat{\sigma}_{a}^{2}=\frac{1}{2n}\sum_{k=1}^{2n}(X_{k}-X_{k-1}-\hat{\mu})^{2}
\end{equation}

\begin{equation}
\hat{\sigma}_{b}^{2}=\frac{1}{2n}\sum_{k=1}^{2n}(X_{2k}-X_{2k-1}-2\hat{\mu})^{2}
\end{equation}
where $\mu$ and $\sigma_{a}^{2}$ correspond to the maximum likelihood
estimators of $\mu$ and $\sigma^{2}$ and $\sigma_{b}^{2}$ correspond
to an estimator of $\sigma^{2}$ that only uses the subset of (n +
1) observations $X_{0},X_{2},X_{4},.,X_{2n}$ and that formally corresponds
to ½ of the two-period variance. \\
 However, using these estimators for the variance has two drawbacks.
The first one is that these are biased estimators, and the second
one is that when a higher period of observations are utilized, the
sample data length decreases.\\

To improve these aspects, Lo and MacKinlay (1988, p. 46) proposed
using the following estimators for variance:

\begin{equation}
\bar{\sigma}_{a}^{2}=\frac{1}{nq-1}\sum_{k=1}^{nq}(X_{k}-X_{k-1}-\hat{\mu})^{2}
\end{equation}

\begin{equation}
\bar{\sigma}_{c}^{2}=\frac{1}{m}\sum_{k=1}^{nq}(X_{k}-X_{k-q}-q\hat{\mu})^{2}
\end{equation}

\begin{equation}
m=q(nq-q+1)(1-\frac{q}{nq})
\end{equation}
where $\bar{\sigma}_{a}^{2}$ is an unbiased estimator of the variance
of a period, $\bar{\sigma}_{c}^{2}$ is an unbiased estimator of the
variance of q periods, and m is the number of observations used in
the variance estimation less one. This turns the variance estimators
into unbiased ones, and reduces the inaccuracy of estimators when
using data with overlaps {[}\citet{lo1988stock}{]}. \\

By using these last expressions, the ratio-variance test (considering
the variance of q periods and q times the variance of one period)
can be calculated as: 
\begin{equation}
M_{r}(q)=\frac{\bar{\sigma_{c}}^{2}}{\bar{\sigma_{a}}^{2}}-1
\end{equation}
where the statistic $M_{r}(q)$ must be zero to satisfy the random
walk null hypothesis. Additionally, $M_{r}(q)$ can be expressed as:
\begin{equation}
\bar{M}_{r}(q)\cong\frac{2(q-2)}{q}\hat{\rho}(1)+\frac{2(q-2)}{q}\hat{\rho}(2)+..+\frac{2}{q}\hat{\rho}(q-1)
\end{equation}
where $\rho(i)$, is the linear autocorrelation of order $i$ of the
time series, defined by: 
\begin{equation}
\hat{\rho}(i)=\frac{nq\sum_{k=i+1}^{nq}(X_{k}-X_{k-1}-\hat{\mu})(X_{k-i}-X_{k-i-1}-\hat{\mu})}{\sum_{k=1}^{nq}(X_{k}-X_{k-1}-\hat{\mu})^{2}}
\end{equation}
Given that there is a growing consensus that the volatility of the
financial series changes over time, it is important to consider this
fact in the inference process for the $M_{r}(q)$ statistic.\\

In order to incorporate these volatility effects, Lo and MacKinlay
(1988, p. 48) used the approach developed by \citet{white1980heteroskedasticity}
and \citet{white1984nonlinear}. This is a rather general approach
that does not require normality in returns and allows temporal nonlinear
auto-dependence. It is assumed that temporal nonlinear autocorrelations
approach to zero when the time interval grows. Therefore, \citet{lo1988stock}
assumed that deviations of returns about its mean follow a process
with short-term time dependencies, as used in \citet{white1980heteroskedasticity}
and \citet{white1984nonlinear}. Explicitly, it is supposed that:
\\
 
\begin{description}
\item [{a)}] For every t, $E(\varepsilon_{t})=0$ and $E(\varepsilon_{t}\varepsilon_{t-\tau})=0$
for some value of $\tau\neq0$ 
\item [{b)}] The ${\varepsilon_{t}}$ is a $\phi$ mixing with $\phi(m)$
coefficients of the size $\frac{r}{2r-1}$ or it is $\alpha$ mixing
with coefficients $\frac{r}{r-1}$, with $r>1$, such that for every
$\tau\geq=0$ there is some $\delta>0$ for which 
\[
E\lvert\varepsilon_{t}\varepsilon_{t-\tau}\lvert^{2(r+\delta)}<\Delta<\infty
\]

\item [{c)}] $\lim\limits _{nq\rightarrow\infty}\frac{1}{nq}\sum_{t=1}^{nq}E(\varepsilon_{t}^{2})=\sigma_{0}^{2}<\infty$ 
\item [{d)}] For every t $E(\varepsilon_{t}\varepsilon_{t-j},\varepsilon_{t}\varepsilon_{t-k})=0$
for any $j\neq k$ \\
 
\end{description}
Under these assumptions, we can calculate the statistical $\bar{M}_{r}(q)$,
which considers asymptotic behavior and in which short-term dependencies
are canceled:\\
 
\begin{equation}
\bar{M}_{r}(q)=\sum_{j=1}^{q-1}\frac{2(q-1)}{q}\hat{\rho}(j)
\end{equation}
Since $\bar{M}_{r}(q)$ is approaching zero under the null hypothesis,
we only need to calculate the value of the asymptotic variance of
this statistic. Considering that this set of hypotheses assumes that
the $\hat{\rho}(j)$ autocorrelations are asymptotically non-correlated,
it is possible to calculate the asymptotic variances of each autocorrelation
$\hat{\rho}(j)$ and therefore the variance of the statistical $\bar{M}_{r}(q)$.
This is because according to expressions (10) and (11), the statistical
$\bar{M}_{r}(q)$ can be considered a weighted average of the autocorrelations
of the first $(q-1)$ lags.\\

Therefore, it is possible to calculate the asymptotic variance of
each autocorrelation $\hat{\rho}(j)$ as: 
\begin{equation}
\hat{\delta}(j)=\frac{nq\sum_{k=j+1}^{nq}(X_{k}-X_{k-1}-\hat{\mu})^{2}(X_{k-j}-X_{k-1-j}-\hat{\mu})^{2}}{[\sum_{k=j+1}^{nq}(X_{k}-X_{k-1}-\hat{\mu})]^{2}}
\end{equation}
Then the asymptotic variance $\hat{\theta}(q)$ of the statistical
$\bar{M}_{r}(q)$ is calculated as: 
\begin{equation}
\hat{\theta}(q)=\sum_{j=1}^{q-1}[\frac{2(q-1)}{q}]^{2}\hat{\delta}(j)
\end{equation}
Considering the asymptotic theory, a statistical z can be calculated
as: 
\begin{equation}
z^{*}(q)=\frac{\sqrt{nq}(\bar{M}_{r}(q))}{\sqrt{\hat{\theta}(q)}}
\end{equation}
and has a normal distribution. Now this statistical $z^{*}(q)$ can
be used to test the null hypothesis. Specifically, if the value of
the statistical $\bar{M}_{r}(q)$ is outside the limits of $[-1.96\sqrt{\hat{\theta}(q)},1.96\sqrt{\hat{\theta}(q)}]$
then the hypothesis of random walk is rejected with a 95\% confidence
level.\\

Before using the random walk test, a descriptive statistical analysis
of the time series will be carried out.

\section{Evolution of result of the variance ratio test}

\subsection{Data Characterization}

The analysis was performed using a weekly data series from January
7, 1950, to March 28, 2013, including 
\begin{itemize}
\item CRSP SP500 market value-weighted; 
\item CRSP NYSE/AMEX/NASDAQ/ARCA weighted by market value; and 
\item CRSP NYSE/AMEX/NASDAQ/ARCA equally weighted. 
\end{itemize}
These indices show the evolution of the weighted average of a significant
part of the shares traded on the New York exchanges. The difference
between indexes depends on two factors:\\
 \\
 1. The relative weight of each stock included in the index; and \\
 2. The type of assets that make up the index. \\
 \\
 The table below shows some basic statistics of two of the indexes
considered.\\

\begin{table}[H]
\begin{centering}
\begin{tabular}{|c|c|c|}
\hline 
Statistic  & CRSP Equally Weighted  & CRSP Value Weighted \tabularnewline
\hline 
Min  & -28.8\%  & -20.24\% \tabularnewline
\hline 
Max  & 14.77\%  & 12.64\% \tabularnewline
\hline 
Mean  & 0.348\%  & 0.198\% \tabularnewline
\hline 
Standard Deviation  & 2.095\%  & 2.080\% \tabularnewline
\hline 
Skewness  & -1.0652  & -0.7826 \tabularnewline
\hline 
Kurtosis  & 12.0012  & 10.1382 \tabularnewline
\hline 
\end{tabular}\protect\caption{Descriptive Statistics Weekly Frequency Data {[}07-01-1950 to 28-03-2014{]} }

\par\end{centering}

\centering{}\label{Tabla 1} 
\end{table}

It was observed that continuously compounded return of the CRSP Equally
Weighted Index had a higher average, higher standard deviation, higher
bias, and higher kurtosis than the CRSP Value-Weighted Index.\\

The figures below show time series, histograms, linear autocorrelations,
and quadratic autocorrelations associated with continuously compounded
returns of the CRSP Equal Weighted and CRSP Value Weighted indexes.\\

\begin{figure}[H]
\centering \includegraphics[width=0.8\linewidth]{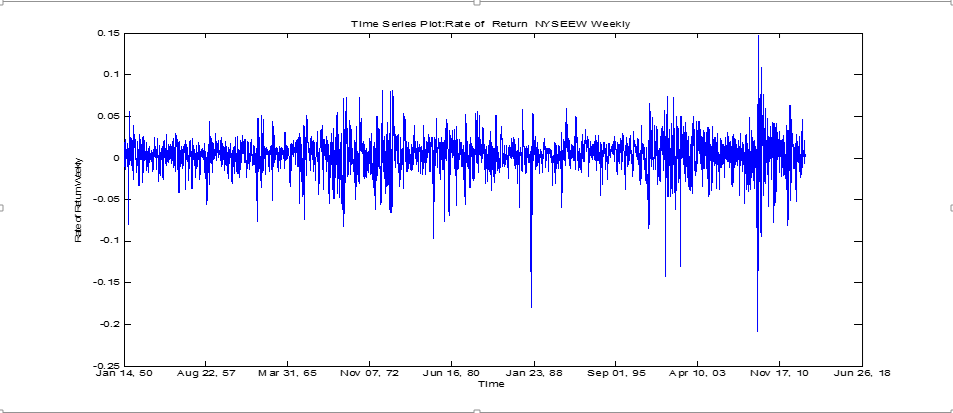} \protect\caption{Series of the weekly rate of returns of CRSP Equal Weighted}

\label{fig:Fig1} 
\end{figure}

\begin{figure}[H]
\centering \includegraphics[width=0.8\linewidth]{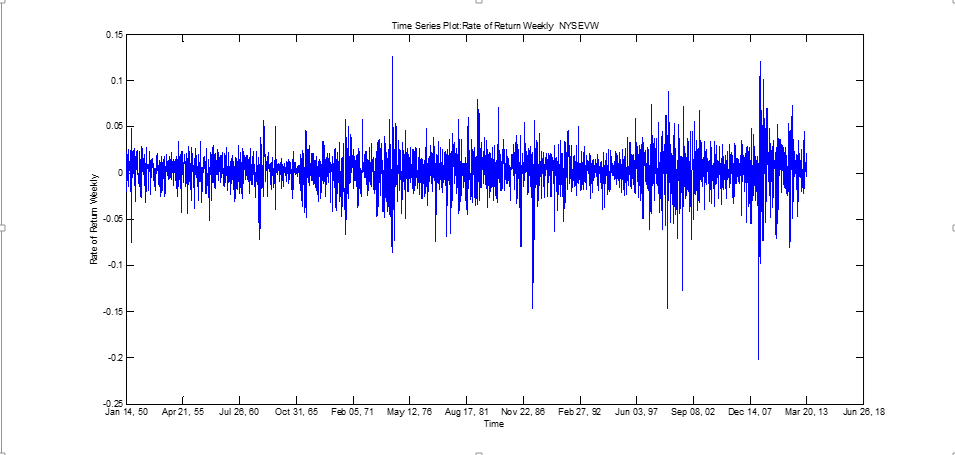} \protect\caption{Series of the weekly rate of returns of CRSP Value Weighted}

\label{fig:Fig2} 
\end{figure}

\begin{figure}[H]
\centering \includegraphics[width=0.8\linewidth]{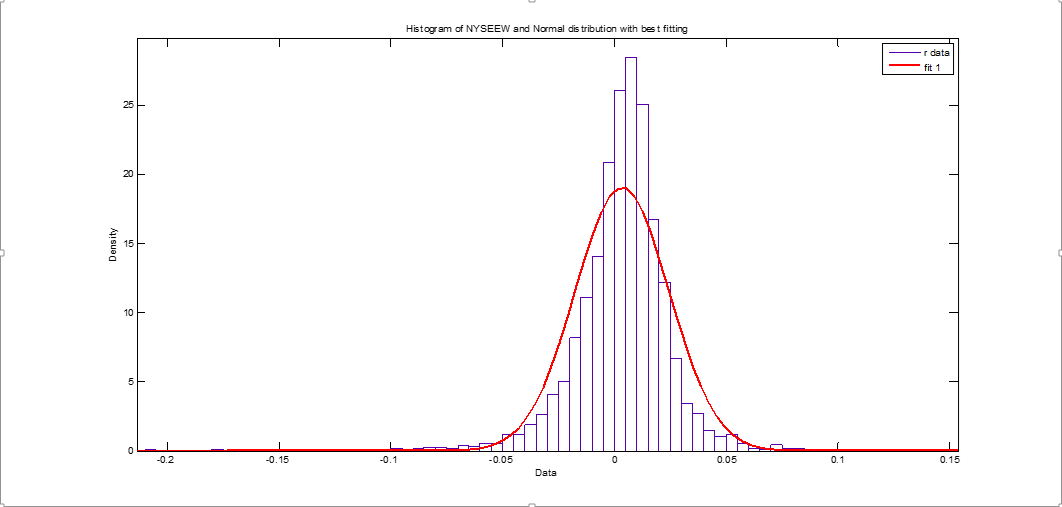} \protect\caption{Histogram of weekly rate of returns of CRSP Equal Weighted}

\label{fig:Fig3} 
\end{figure}

\begin{figure}[H]
\centering \includegraphics[width=0.8\linewidth]{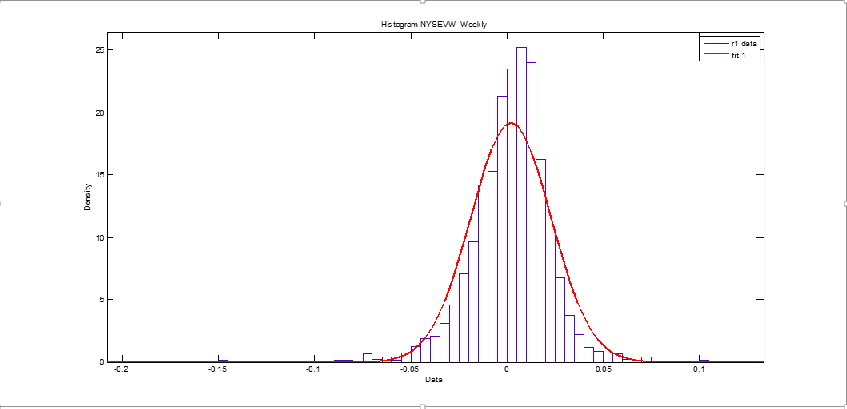} \protect\caption{Histogram of weekly rate of returns of CRSP Value Weighted}

\label{fig:Fig4} 
\end{figure}

\begin{figure}[H]
\centering \includegraphics[width=0.8\linewidth]{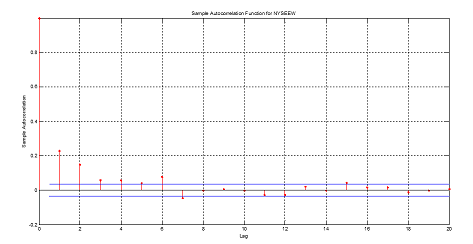} \protect\caption{Linear autocorrelation of rate of returns of CRSP Equal Weighted}

\label{fig:Fig5} 
\end{figure}

\begin{figure}[H]
\centering \includegraphics[width=0.9\linewidth]{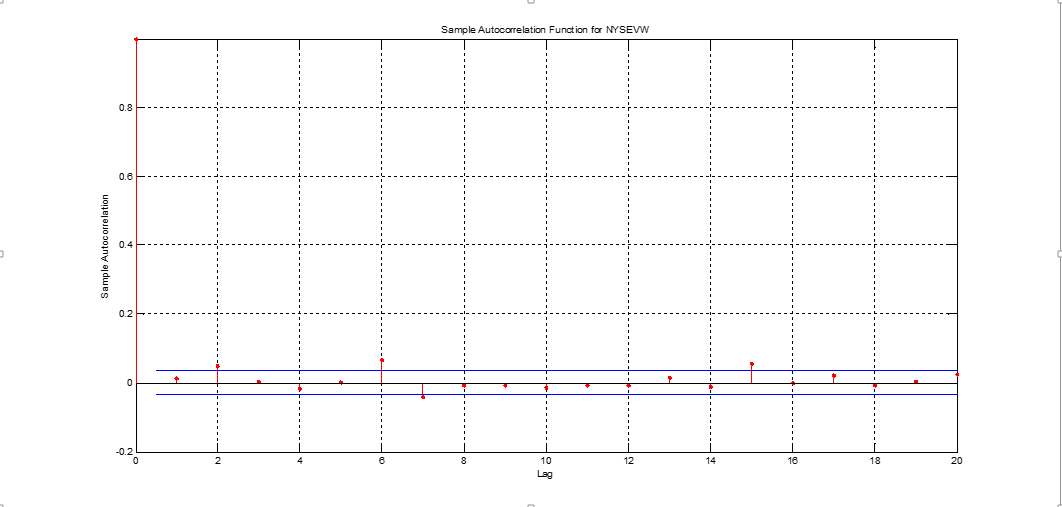} \protect\caption{Linear autocorrelation of rate of returns of CRSP Value Weighted}

\label{fig:Fig6} 
\end{figure}

\begin{figure}[H]
\centering \includegraphics[width=0.8\linewidth]{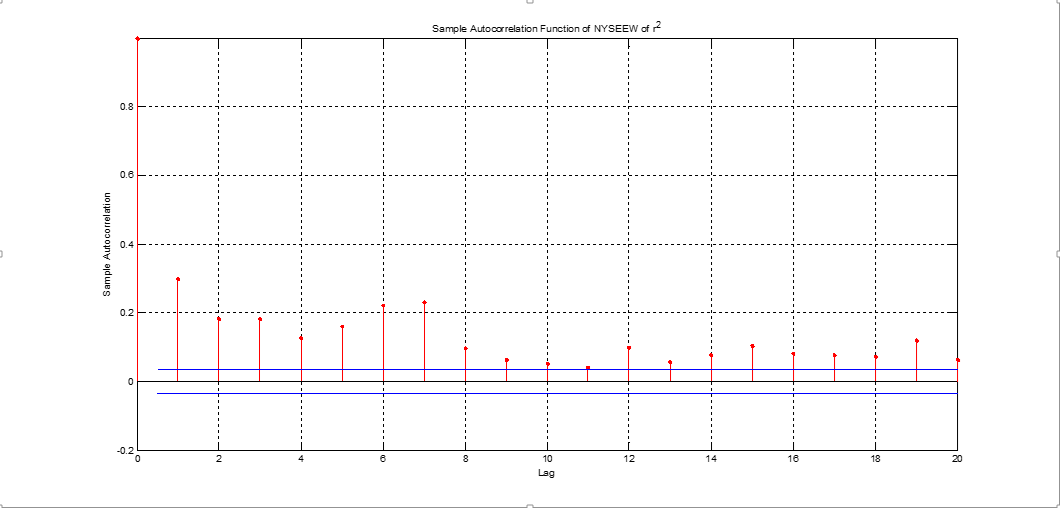} \protect\caption{Autocorrelation of square of rate of returns of CRSP Equal Weighted}

\label{fig:Fig7} 
\end{figure}

\begin{figure}[H]
\centering \includegraphics[width=0.8\linewidth]{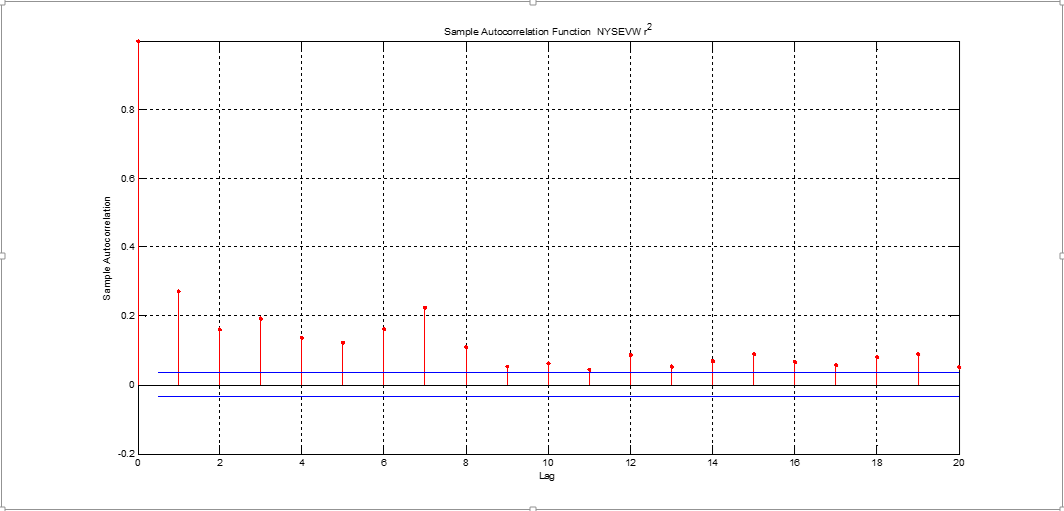} \protect\caption{Autocorrelation of square of rate of returns of NYSE Equal Weighted}

\label{fig:Fig8} 
\end{figure}

From these figures, it is possible to deduce that:
\begin{description}
\item [{a)}] The returns of both indexes have time-dependent volatility
variations (Figures 1 and 2); 
\item [{b)}] The empirical distribution of returns moves away from the
normal distribution and shows high levels of leptokurtosis (Figures
3 and 4); 
\item [{c)}] The linear autocorrelations for the CRSP NYSE AMEX NASDAQ
ARCA Equally Weighted index show non-zero values at the first time
lags (Figure 5) 
\item [{d)}] The linear autocorrelations of the CRSP NYSE AMEX NASDAQ ARCA
Value Weighted are negligible for all lags (Figure 6); and 
\item [{f)}] The autocorrelations of the square of the rate of return are
higher than zero for both index (Figures 7 and 8). It is possible
to estimate that these autocorrelations are still significantly different
from zero after 20 lags. 
\item [{g)}] The above-mentioned issues are consistent with a stochastic
process in which the variance changes in time with some degree of
predictability.
\end{description}

\subsection{Analysis of the results}

The results of applying the variance ratio test to the indices: CRSP
SP500, CRSP NYSE/AMEX/NASDAQ/ARCA Value Weighted and CRSP NYSE AMEX
NASDAQ ARCA Equally Weighted are shown below. The variance ratios
were calculated using periods of 2, 4, 8, and 16 weeks for weekly
data ranging from July 1, 1950 to March 28, 2013.

\begin{table}[h!]
\begin{centering}
\protect\caption{Variance ratio test for SP500}

\par\end{centering}

\centering{}\label{tab:table2} %
\begin{tabular}{lccccccr}
q  & 2  & 4  & 8  & 16  &  &  & \tabularnewline
\hline 
ratio  & 0.9795  & 1.0067  & 1.0082  & 0.9975  &  &  & \tabularnewline
Statistic Z(q)  & -0.6635  & 0.1217  & 0.0993  & -0.0212  &  &  & \tabularnewline
Critical value  & (+/-)1.96  & (+/-)1.96  & (+/-)1.96  & (+/-)1.96  &  &  & \tabularnewline
p-value  & 0.507  & 0.9031  & 0.9209  & 0.9831  &  &  & \tabularnewline
\end{tabular}
\end{table}

\begin{table}[h!]
\begin{centering}
\protect\caption{Variance ratio test for CRSP Value Weighted}

\par\end{centering}

\centering{}\label{tab:table3} %
\begin{tabular}{lccccccr}
q  & 2  & 4  & 8  & 16  &  &  & \tabularnewline
\hline 
ratio  & 1.0132  & 1.0716  & 1.115  & 1.1202  &  &  & \tabularnewline
Statistic Z(q)  & 0.4053  & 1.2321  & 1.3199  & 0.98  &  &  & \tabularnewline
Critical value  & (+/-)1.96  & (+/-)1.96  & (+/-)1.96  & (+/-)1.96  &  &  & \tabularnewline
p-value  & 0.6853  & 0.2179  & 0.1869  & 0.3271  &  &  & \tabularnewline
\end{tabular}
\end{table}

\begin{table}[H]
\begin{centering}
\protect\caption{Variance ratio test for CRSP Equal Weighted}

\par\end{centering}

\centering{}\label{tab:table4} %
\begin{tabular}{lccccccr}
q  & 2  & 4  & 8  & 16  &  &  & \tabularnewline
\hline 
ratio  & 1.2295  & 1.5243  & 1.8186  & 1.9693  &  &  & \tabularnewline
Statistic Z(q)  & 6.305  & 8.1302  & 8.5506  & 7.2269  &  &  & \tabularnewline
Critical value  & (+/-)1.96  & (+/-)1.96  & (+/-)1.96  & (+/-)1.96  &  &  & \tabularnewline
p-value  & 0.000  & 0.000  & 0.000  & 0.000  &  &  & \tabularnewline
\end{tabular}
\end{table}

\noindent The results shown in tables 2, 3, and 4 show that: 
\begin{itemize}
\item The null hypothesis of the random walk is not rejected for the SP500
and CRSP Value-Weighted indexes (both weighted by market value) with
a significance level of 5\%. 
\item The null hypothesis is rejected for the CRSP Equally Weighted index. 
\end{itemize}
These results differ from those reported in Lo and MacKinlay (1988,
pp. 52-54). They indicated that the null hypothesis was rejected for
the CRSP Value-Weighted index and the CRSP Equal-Weighted index. This
happened because they used weekly sample data from June 9, 1962 up
only to December 26, 1985. As will be shown below, until the 1980s,
the null hypothesis, in effect, was rejected for both indices. But,
after the 1990s, the null hypothesis was not rejected for the CRSP
Value-Weighted index.\\

The evolution of the behavior of the various indices was analyzed
using weekly sample data, from July 1, 1950, until March 28, 2013;
that is, 1,601 weekly windows (subsamples).\\

The first window began on July 1, 1950, and from that date, the following
1,601 weeks were considered. Then the p-value of the ratio-variance
test was calculated for each subsample. After that, the window moved
a week (the second subsample began on January 14, 1950) and the p-value
of the test was recalculated, and so on until the last window in the
sample.\\

In the following graphs, the p-values obtained for each of the 1,699
subsamples are reported. The vertical axis indicates the p-value for
each subsample and the horizontal one indicates the last date of each
subsample. We must remember that the null hypothesis is rejected in
the case when the p-value is less than 0.05. In this way, we can see
the evolution in the behavior of each index.\\

\begin{figure}[H]
\centering \includegraphics[width=0.9\linewidth]{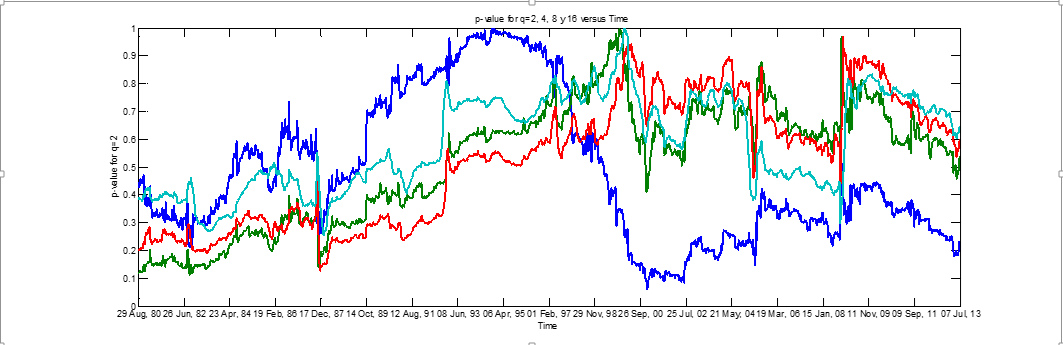} \protect\caption{Evolution of p-value of VR test {[}SP500{]}}

\label{fig:Fig9} 
\end{figure}

\begin{figure}[H]
\centering \includegraphics[width=0.8\linewidth]{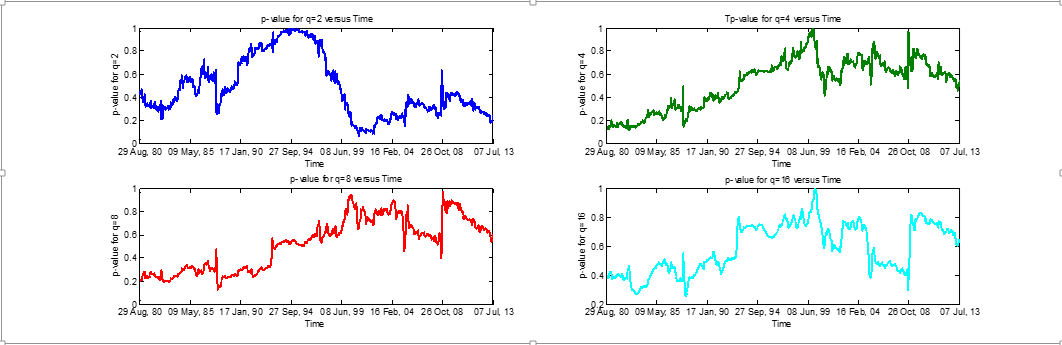} \protect\caption{Evolution of p-value of VR test {[}SP500{]}}

\label{fig:Fig10} 
\end{figure}

Figures 9 and 10 show the evolution of the performance of the SP500
index. Figure 9 displays the evolution of the p-value for q values
equal to 2, 4, 8, and 16. The blue line in Graph 9 corresponds to
q = 2, the green one to q = 4, the red one to q = 8, and the light
blue one to q = 16. Each figure shows that the p-value of each series
is never less than 5\%, and therefore the null hypothesis was never
rejected for any of the windows considered.

\begin{figure}[H]
\centering \includegraphics[width=0.8\linewidth]{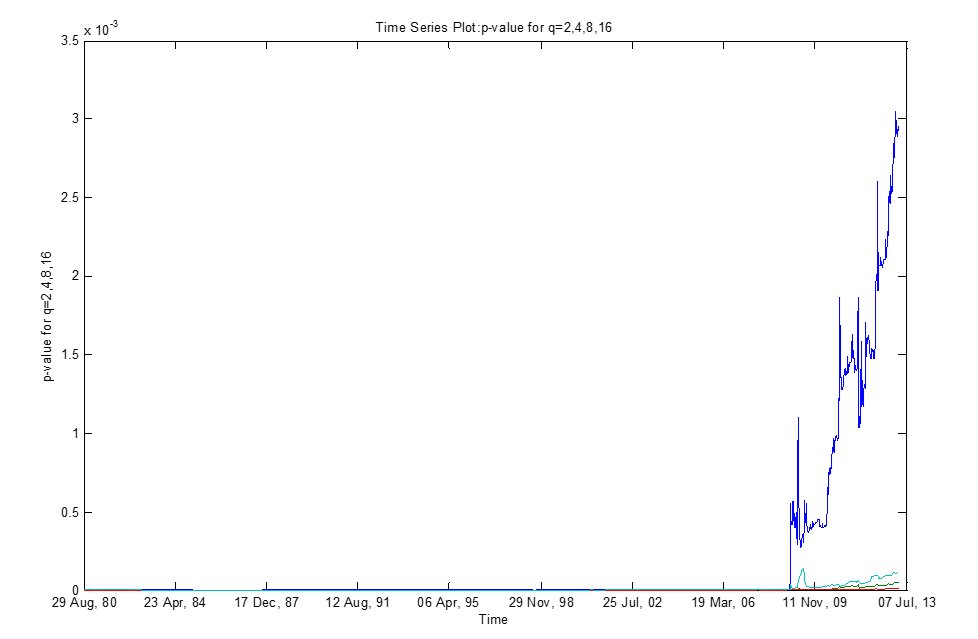} \protect\caption{Evolution of p-value of VR test {[}CRSP NYSE/AMEX/NASDAQ/ARCA Equal
Weighted{]}}

\label{fig:Fig11} 
\end{figure}

\begin{figure}[H]
\centering \includegraphics[width=0.8\linewidth]{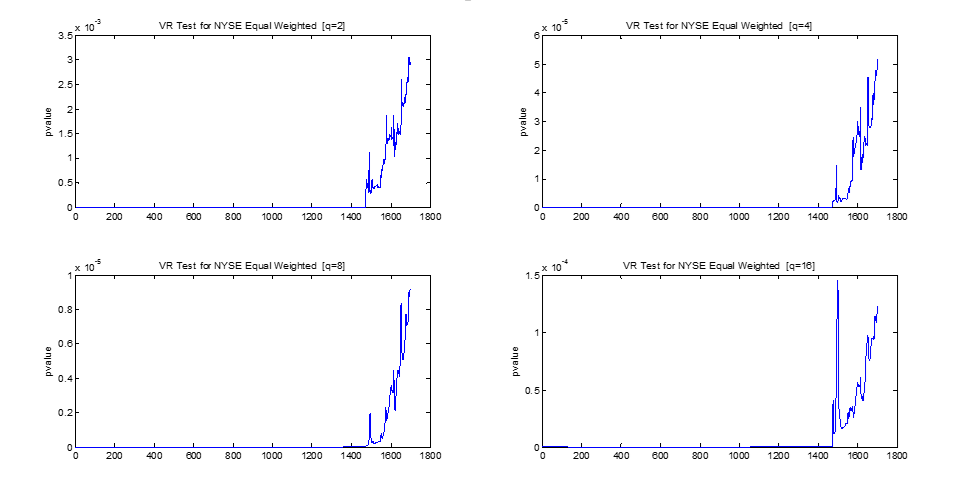} \protect\caption{Evolution of p-value of VR test {[}CRSP NYSE/AMEX/NASDAQ/ARCA Equal
Weighted{]}}

\label{fig:Fig12} 
\end{figure}

Figures 11 and 12 show the evolution in the behavior of the CRSP Equal-Weighted
index. Figure 11 shows the evolution of the p-value for values of
q equal to 2, 4, 8, and 16. Both figures show that the p-value of
each series is less than 5\% for all subsamples. These figures show
that the null hypothesis is rejected in each of the windows considered.

\begin{figure}[H]
\centering \includegraphics[width=0.8\linewidth]{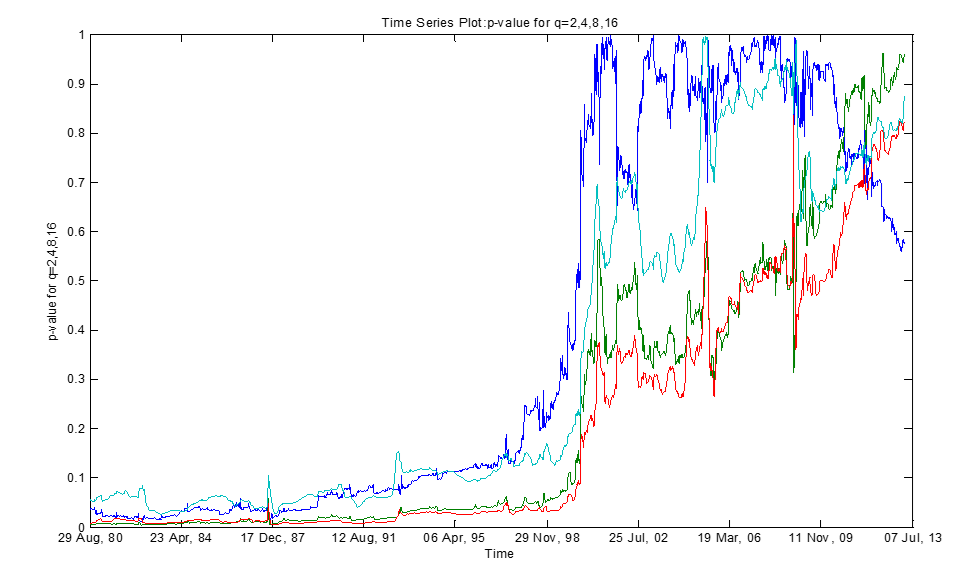} \protect\caption{Evolution of p-value of VR test {[}CRSP NYSE/AMEX/NASDAQ/ARCA NYSE/AMEX/NASDAQ/ARCA
Value Weighted{]}}

\label{fig:Fig13} 
\end{figure}

\begin{figure}[H]
\centering \includegraphics[width=0.8\linewidth]{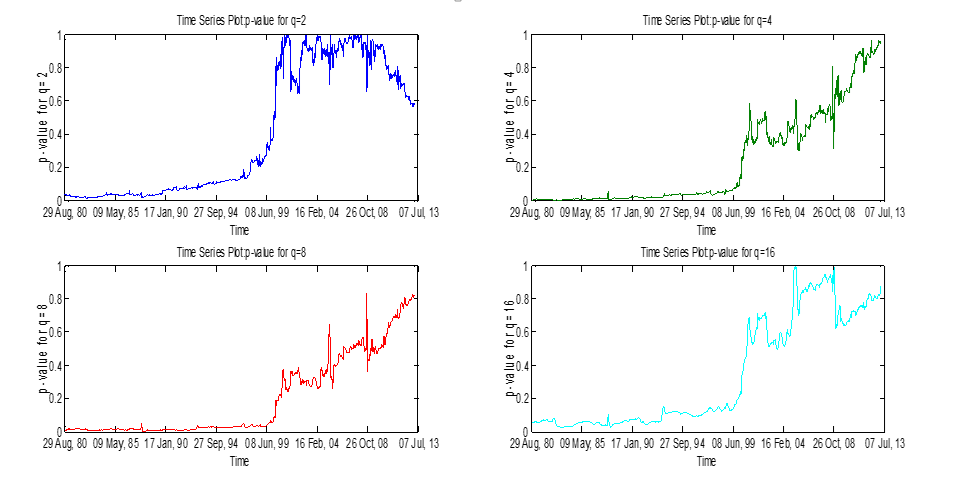} \protect\caption{Evolution of p-value of VR test {[}CRSP NYSE/AMEX/NASDAQ/ARCA Value
Weighted{]}}

\label{fig:Fig14} 
\end{figure}

Figures 13 and 14 show the behavior of the CRSP NYSE Value-Weighted
index. These figures display the evolution of the p-value for values
of q equal to 2, 4, 8, and 16. Both figures show that the p-value
of each series is less than 5\% until the 1980s. However, from the
end of the 1990s, the null hypothesis is not rejected for all of the
considered windows. \\

Thus, clearly the various index exhibit different behaviors from the
end of the 1990s. As we have seen, the SP500 and the CRSP NYSE Value-Weighted
are weighted by market value of the shares that constitute them. In
this case, a higher weighting is given to stocks of large corporations
and a low weighting to smaller ones. \\

On the other hand, for the CRSP NYSE Equal-Weighted index, the same
assets of the CRSP Value-Weighted are included, but this time with
equal weights. For this reason, the different behavior of these two
index starting from the 1990s has our attention. \\

To explain these divergent behaviors, various factors can be postulated,
such as: 
\begin{description}
\item [{i)}] Different behaviors between the assets of lower market value
and the assets of companies of greater market value; 
\item [{ii)}] A change in the behavior of assets with higher market value; 
\item [{iii)}] Problems with the power of the ratio-variance test when
using finite samples and overlapping data; and 
\item [{iv)}] Returns generating processes with different characteristics
to those assumed in the null hypothesis test, among other factors. 
\end{description}
It is also possible that the cause of this different behavior since
the 1990s is due to problems with the statistical test used. To determine
if this was the case, another test that is statistically more robust
for finite size samples and overlapping data will be used.\\

Although the implementation of the ratio-variance analysis is relatively
straightforward, performing the statistical inference process is not
so simple. One of the major complications for this test is the fact
that the inference uses: 
\begin{description}
\item [{a)}] An asymptotic distribution of the statistic; and 
\item [{b)}] Overlapping data to calculate variances for periods of time
exceeding a period. 
\end{description}
As we saw above, the use of overlapping data was suggested by Lo and
MacKinlay (1988, p. 46) to improve the power of the test. However,
overlapping data produces difficulties for the inference analysis
{[}\citet{richardson1991tests}{]}. In addition, the sample sizes
used were finite. This had the implication that the empirical distribution
of the statistic may have been quite far away from the asymptotic
distribution.\\

The variance ratio test proposed by \citet{kim2006wild} is a robust
test for finite sample sizes and overlapping data. This test uses
the bootstrapping method to estimate the empirical distribution {[}\citet{efron1979bootstrap}{]}.
Therefore, it makes an inference using the empirical distribution
of the statistic and not asymptotic distribution. The consequences
of applying this last robust test to the CRSP Value-Weighted indices
and the CRSP Equal-Weighted are shown:

\begin{figure}[H]
\centering \includegraphics[width=0.8\linewidth]{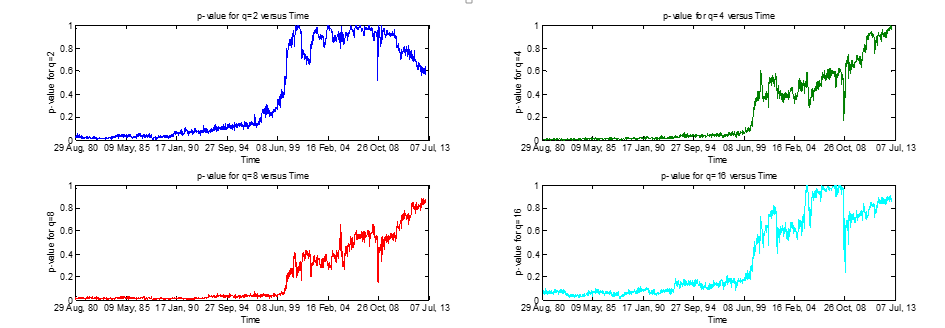} \protect\caption{Evolution of p-value of VR test {[}NYSE Value Weighted{]}}

\label{fig:Fig15} 
\end{figure}

\begin{figure}[H]
\centering \includegraphics[width=0.8\linewidth]{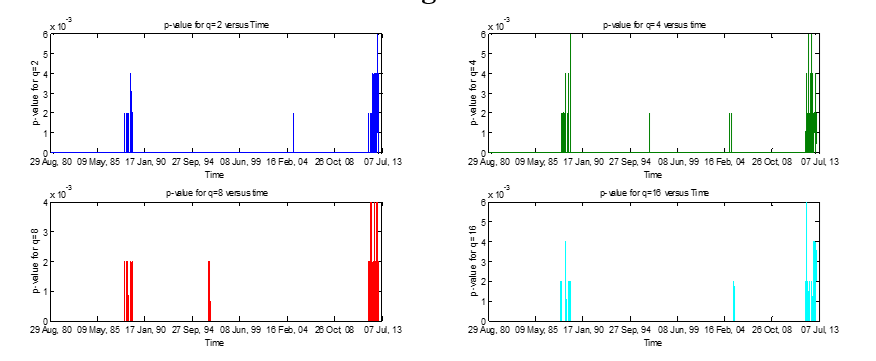} \protect\caption{Evolution of p-value of VR test {[}NYSE Equal Weighted{]}}

\label{fig:Fig16} 
\end{figure}

Charts 15 and 16 show the results of applying the ratio-variance test
by \citet{kim2006wild} to CRSP NYSE Equal-Weighted and CRSP NYSE
Value-Weighted indexes.\\

These findings are similar to those mentioned above. Charts 15 and
16 show the same behavior again: 
\begin{description}
\item [{a)}] The null hypothesis is rejected in all windows for the CRSP
NYSE Equal-Weighted Index. 
\item [{b)}] It is not rejected for index CRSP NYSE Value-Weighted starting
from the end of the end of the 1990s. 
\end{description}
Thus, these results confirm those obtained with the initially applied
test, and therefore the difference between the indexes behaviors is
not a problem of the statistical test and must be explained by other
factors.

\subsection{Summing up the historical evolution of results of variance ratio
test}

The question of whether the financial asset prices come after a random
walk or are a martingale-type sequence has strong implications on
the efficiency of marketplaces.\\

If the difference in the logarithm of prices for financial assets
follows a random walk, then returns are not predictable, and investors
are unable to obtain higher profits consistently over time. \\

Critical studies carried out considering the information up to the
1980s allowed researchers to establish that the main index of the
New York Stock Exchange did not follow a random walk. The null hypothesis
is rejected for two of the major stock index. \\

However, when this study is carried out with a much wider sample,
these conclusions can be refuted. The outcomes of using the ratio
variance test on 1,699 1600 rolling windows (each composed of 1,601
weeks) permit us to express the following:
\begin{description}
\item [{a)}] The null hypothesis is rejected for each sub-sample of the
equally-weighted index (NYSEEW). 
\item [{b)}] It is never rejected for the SP 500 index. 
\item [{c)}] For the Value-weighted index (NYSEVW), the null hypothesis
is rejected until the first part of the 1990s. But it cannot be refuted
from that epoch to the present time. 
\end{description}
The fact that the random walk null hypothesis is not refuted for the
CRSP Value-Weighted index from the end of the 1990s can have important
connotations and greater significance for investors. In fact, from
the point of view of traditional financial theories, this index is
commonly regarded as an estimator of the market portfolio and therefore
the manner in which investors should place their money in diversified
portfolios. \\

On the other hand, the CRSP Equal-Weighted index represents the evolution
of a portfolio in which the same amount of money has been put in each
component stock. This final portfolio is in theory seen as an inefficient
portfolio.\\

These results could endorse the thesis that, from the 1990s, the New
York Stock Exchange market followed a random walk. Apparently, a structural
change took place in this period in the NYSE.

\section{A model of non-synchronous trading and portfolio effects}

As shown above, the variance ratio test gives a different result when
applied to the equally weighted index (the null hypothesis is rejected)
from the index weighted by market value (the null hypothesis is not
rejected).\\

This research offers a simple model that could explain the differences
in behavior of the equally weighted index and index weighted by value.
For this purpose, non-trading, non-synchronous trading, and portfolio
effects will be considered. \\
 \\
 The idea of the proposed approach is to study these phenomena considering
that the problem of inference must be performed:
\begin{description}
\item [{a)}] With a finite size sample data, 
\item [{b)}] With a portfolio with finite asset number, and 
\item [{c)}] Without using any model of equilibrium prices. 
\end{description}
It is assumed that the set of individual financial assets follows
multivariate normal process. The non-trading effect is introduced
to generate individual returns, and then its impact on the variance
ratio test is analyzed. The implementation of this approach is explained
below.\\

To capture the non-synchronous trading effect, we used a similar approach
to the one presented by \citealp{lo1990econometric}, with the difference
that asymptotic behaviors were not considered, but instead the actual
problem of inference, was addressed using finite size samples. \\

For this, consider a set of $N$ assets with unobservable virtual
continuously compounded returns $r_{it}$ \ $i=1,2,..n$ at time
$t$. It is assumed that the virtual returns are generated by the
following vector stochastic process: 
\begin{equation}
r_{t}=\mu+\varepsilon_{t}
\end{equation}
where $\mu$ is the mean returns vector and $\varepsilon_{t}$ is
constructed from $n$ independent and identically distributed random
variables with zero mean and covariance matrix $V$. These virtual
returns would be observed if the asset was regularly traded on each
of the registration periods. It is assumed that, in each period $t$,
the asset i is traded with constant probability $\pi_{i}$. \\
 \\
 For each asset $i$ in the period $t$, there is an observed return
$r_{i(t)}^{^{o}}$, which depends on:
\begin{description}
\item [{a)}] The virtual returns, and 
\item [{b)}] Whether the asset is traded in each period. 
\end{description}
\vspace{0.2cm}
 The observed return $r_{i(t)}^{^{o}}$ is defined in the follow way:
\\
 
\[
\begin{array}{lcl}
\mbox{If the asset is not traded} & \Rightarrow & r_{i(t)}^{^{o}}\mbox{ is zero}\\
\mbox{If the asset is traded} & \Rightarrow & r_{i(t)}^{^{o}}\ \mbox{is the sum of previous returns not traded}
\end{array}
\]
\\
 For example, if the asset $"i"$ was not traded in the last three
periods, the observed return $r_{i(t)}^{^{o}}$ would be: 
\begin{description}
\item [{i)}] $r_{i(t-3)}+r_{i(t-2)}+r_{i(t-1)}+r_{i(t)}$ \ if the asset
was traded 
\item [{ii)}] equal to zero if was not traded 
\end{description}
\vspace{0.3cm}
 On the other hand, the virtual price of a financial asset can be
represented as: 
\begin{equation}
P_{i}(t)=P_{i}(0)\exp(r_{i}(1)+r_{i}(2)+\cdots+r_{i}(t))
\end{equation}
and then the accumulated return in t periods shall be equal to: 
\begin{equation}
R_{i}(t)=\ln(P_{i}(t))-\ln(P_{i}(0))=r_{1}+r_{2}+\cdots+r_{t}
\end{equation}
In this form, the virtual price depends on the sum of all the past
returns. However, when the financial asset is not traded, the observed
price differs from the virtual price. The observed price is defined
by: 
\begin{equation}
P_{i}^{o}(t)=\left\lbrace \begin{array}{ll}
P_{i}(t) & \mbox{when asset is traded}\\
P_{i}^{o}(t) & \mbox{when the asset is not traded}
\end{array}\right.
\end{equation}
and then, by (16) and (17) the accumulated observed return of an asset
$i$ until the instant $t$ is equal to: 
\begin{equation}
R_{i}^{o}(t)=\left\lbrace \begin{array}{ll}
R_{i}(t) & \mbox{when asset is traded with probability}\ \pi_{i}\\
R_{i}^{o}(t) & \mbox{when asset is not traded with probability}\ \pi_{i}
\end{array}\right.
\end{equation}
From a practical view, the original definition of observed returns
$r_{i}^{o}(t)$ is difficult to implement. But one can see that the
observed returns and the accumulated observed returns are related
by: 
\begin{equation}
r_{i}^{o}(t)=R_{i}^{o}(t)-R_{i}^{o}(t-1)
\end{equation}
This last expression is a friendlier way to realize the simulation
process, and thus it will be used in the following analysis. \\

The process used to generate samples of time series of observed returns
is the following: 
\begin{description}
\item [{i)}] A time series sample of virtual returns $r_{i}(t)$ is generated
using a multivariate normal process N($\mu$,V). 
\item [{ii)}] Using this sample, another sample of accumulated virtual
returns $R_{i}(t)$ is generated. 
\item [{iii)}] Using Equation (19), the accumulated time series of observed
returns $R_{i}^{o}(t)$ is generated. 
\item [{iv)}] Then, with the accumulated observed returns series, a time
series of observed returns $r_{i}^{o}(t)$ is generated using equation
(20). 
\end{description}
\vspace{0.2cm}
 The above algorithm is used to introduce the effect of non-synchronous
trading on the virtual returns time series. Note that this model could
be used to study other subjects, such as the incidence of non-synchronous
trading over the selection of portfolios and parameter estimation,
among others. In our case an equilibrium price model is not assumed
as in \citet{atchison1987nonsynchronous}, \citet{lo1990econometric},
and \citet{boudoukh1994tale}. Our model allows the study of problems
considering finite size samples, without assuming asymptotic behaviors.
\\

To appreciate the joint effect that non-synchronous trading and the
portfolio have on the inference of the variance ratio test when the
sample size is finite, several different two asset portfolios were
built. Non-trading was introduced to only one of the individual assets.
Later on, an estimation of the correlation in weekly profits induced
by non-synchronous trading was undertaken, and the same results were
obtained for a six-asset portfolio with heterogeneous features similar
to those used in \citet{boudoukh1994tale}.

\section{Main results of the effects of trading and portfolio on the VR test}

\subsection{Case with two financial assets}

Estimates are made by means of the Monte Carlo simulation using MATLAB.
Samples of normal multivariate two-asset return series (of 5000 periods
long) were generated. The nontrading effect was applied only to the
second financial asset return series ($r_{2}^{o}(t)$). The nontrading
probability was set at zero for Asset 1 and 20 $\%$ for Asset 2.\\
 Samples of series of indexes returns were built according to: 
\begin{equation}
r_{p}^{o}(t)=(1-\alpha)r_{1}^{o}(t)+\alpha r_{2}^{o}(t)\hspace{1cm}t=1,2,..,5000
\end{equation}
where $\alpha$ was the percentage of wealth assigned to the asset
with nontrading effect. The $\alpha$ parameter can take the values:
$0,0.01,0.02\cdots,1$.\\
 We assumed that the mean of the vector returns was $\mu$={[}0.005,0.005{]},
the standard deviation of asset 1 was 0.005, and the standard deviation
of asset 2 was 0.02. The correlation factor $\rho$ can take values
between -1 and 1 and can change between simulations. We did this to
see what influence the correlation had on the variance ratio test
outcomes.\\

The simulation process goes as follows: 
\begin{description}
\item [{i)}] The correlation parameter $\rho$ took an initial value and
would remain fixed, and for each of its values 10,000 multivariate
time series simulations were generated. 
\item [{ii)}] Each of these time series incorporated the non-synchronous
trading effect. 
\item [{iii)}] With these multivariate time series, one can construct a
portfolio set by means of equation (20), so one has 10.000 different
possible portfolio return series for each value of the $\alpha$ parameter. 
\item [{iv)}] The variance ratio test was applied keeping both $\alpha$
and $\rho$ parameters fixed. The p-value was calculated for the 10.000
possibilities. 
\item [{v)}] If the p-value is less than $10\%$, the null hypothesis of
the random walk is rejected. 
\end{description}
After that, the reject frequency was plotted versus the $\rho$ and
$\alpha$ parameter values, as illustrated in Figure 17. It shows
the frequency percentage at which the null hypothesis was rejected
using the variance ratio test. This graphic shows that when the non-trading
effect was introduced, the variance ratio test rejected the null hypothesis
to a higher degree than when non-trading was not present.\\

It is possible to see in figure $17$ that the rejection of the null
hypothesis was very strong when the correlation was negative. There
was practically no rejection when the correlation was close to zero,
and there was a significant increase of the rejection, as the correlation
got closer to 1.

\begin{figure}[H]
\centering \includegraphics[width=0.8\linewidth]{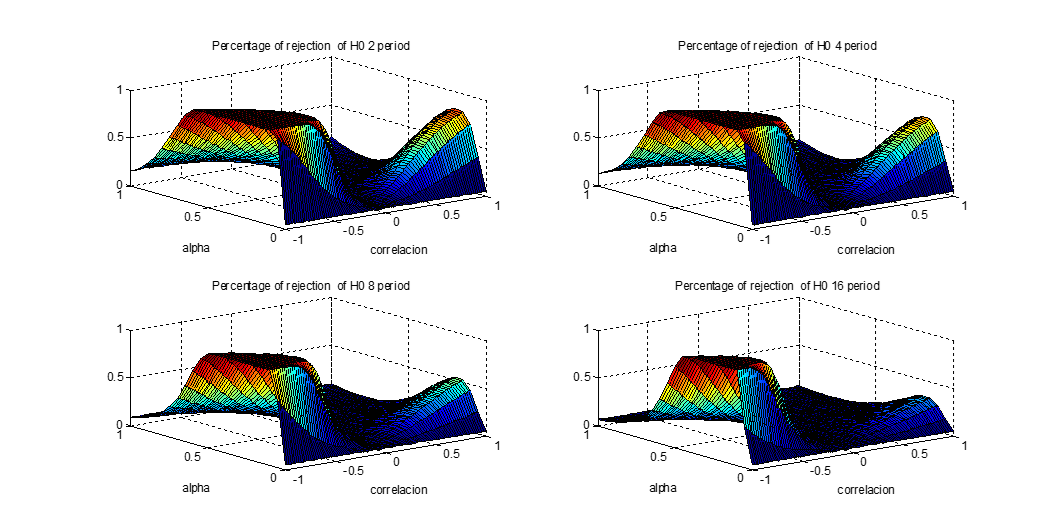} \protect\caption{Portfolio and nontrading effect}

\label{fig:Fig17} 
\end{figure}

Figure 18 shows the frequency percentage at which the null hypothesis
was rejected without non-trading effect. Note that when the time series
did not consider the non-trading effect, the average value of rejection
of the null hypothesis was closer to $10\%$. More importantly, its
value did not depend on the correlation or how to portfolio was built
(the $\alpha$ parameter).

\begin{figure}[H]
\centering \includegraphics[width=0.8\linewidth]{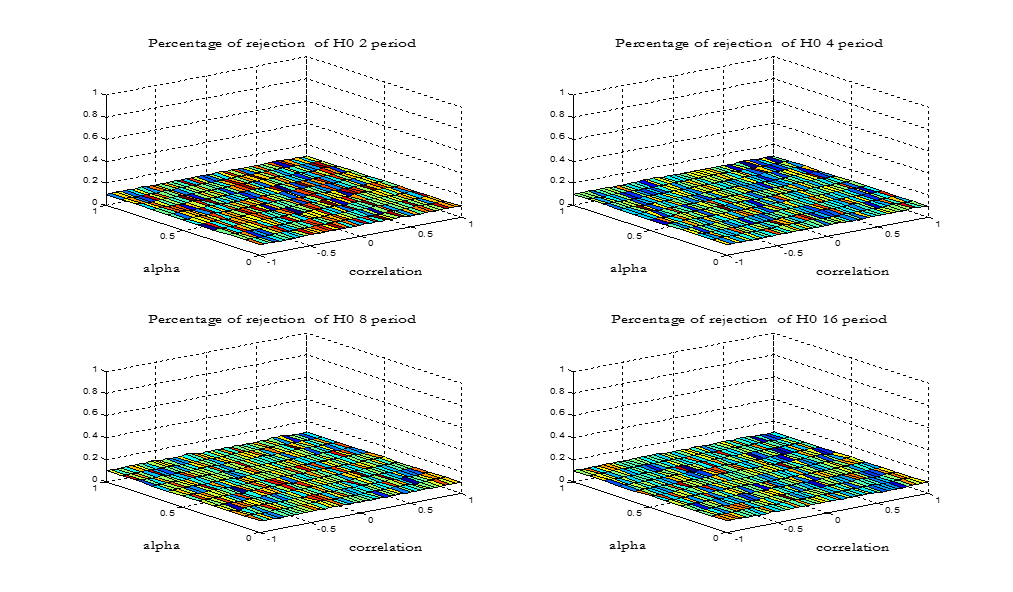} \protect\caption{Percentage of Rejection with Portfolio effect only}

\label{fig:Fig18} 
\end{figure}

The sensitivity of the variance ratio test was very strong to the
portfolio and non-trading effects for different correlation values
between assets. The only case in which its impacts were insignificant
was when the correlation between assets was close to zero.\\

Given that usually the correlation between financial assets is close
to 1, from here we focus on analyzing the portfolio and non-trading
effect for a case in which correlation between assets is equal to
$90\%$. The case is considered representative of the current situation.\\

Figure 19 shows the results obtained when restricting the correlation
value between the assets to a value equal to $90\%$. The graphics
show the percentage that the null hypothesis for the variance ratio
test was rejected. The blue line represents the case of $k=2$, the
green line is the same but for $k=4$, the red line is for $k=8$,
and finally the magenta line is for $k=16$ periods.

\begin{figure}[H]
\centering \includegraphics[width=0.8\linewidth]{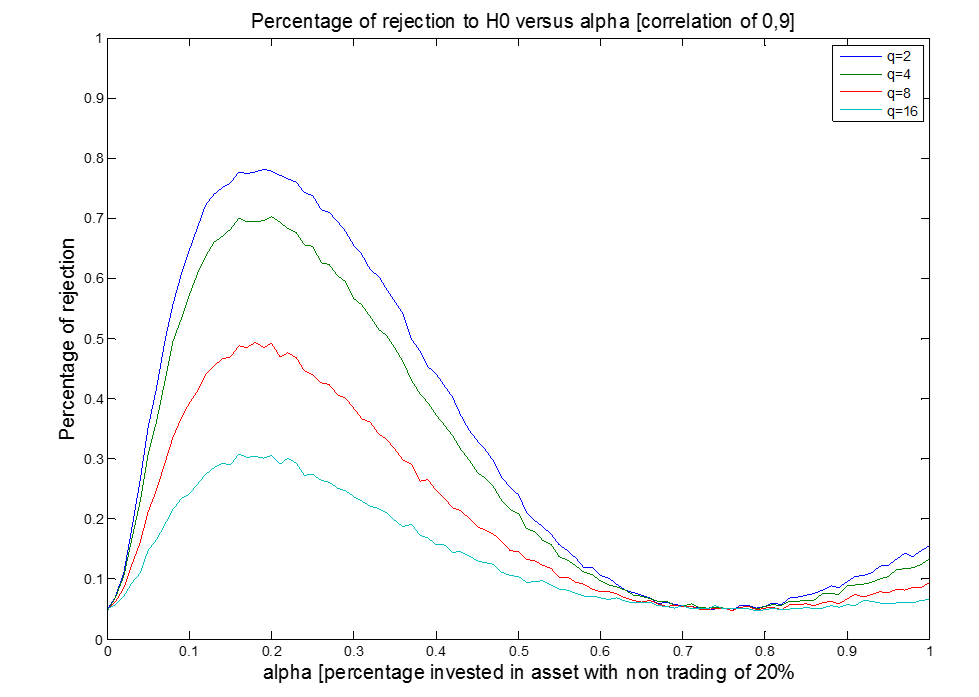} \protect\caption{Portfolio effect for correlation equal 90$\%$}

\label{fig:Fig19} 
\end{figure}

Figure 19 shows that the maximum percentage of null hypothesis rejection
was produced close to the 0.16 for the $\alpha$ value. This rejection
was greater for the variance-ratio two-period test. Reasons for rejection
for this $\alpha$ value are due to the positive autocorrelation.
The rejection percentage value of the null hypothesis was around 70$\%$
for $\alpha$ = 0.16.\\

Figure 20 shows the histogram of the variance-ratio value for $\alpha=0.16$.
The histogram illustrates how the statistical values of variance ratio
of the 10.000 simulated trajectories were distributed.

\begin{figure}[H]
\centering \includegraphics[width=0.7\linewidth]{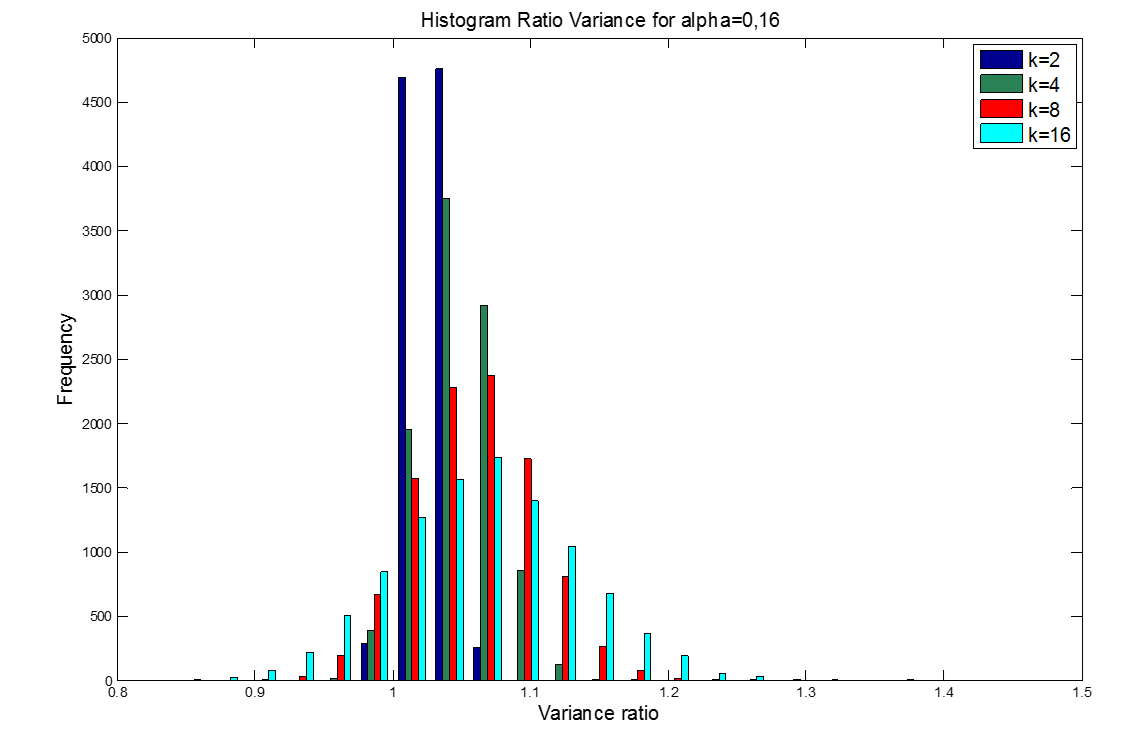} \protect\caption{Histogram for Ratio variance for alpha = 0.16}

\label{fig:Fig20} 
\end{figure}

Traditional models such as \citet{lo1990econometric} estimated the
theoretical impact that non-synchronous trading had on the inference
problem. These authors used the asymptotic behavior of the variance
ratio test statistic. Nevertheless, in this article, the complete
statistical distribution of the variance-ratio values was obtained.
This approach is more powerful than traditional ones because we can
obtain an average value for the variance ratio test and its frequency
distribution.\\

In fact, through simulations, one can visualize the different possible
frequency distributions of the effect of non-synchronous trading for:
\begin{description}
\item [{a)}] Different asset correlations (one can change the $\rho$ values);
and 
\item [{b)}] Different portfolio structures (one can change the $\alpha$
values). 
\end{description}
For example, in Figure 20, the average ratio-variance value for $k=8$
was 1.07. But, Figure 20 also shows that in approximately 25$\%$
of instances, the variance ratio was larger than 1.1. This last piece
of information is very relevant to the statistical analysis because
it is possible in practice (when a single time series is used) that
the value of the observed variance-ratio statistic will be greater
than its expected value. This last fact can lead to incorrect inferences.\\

In short we have found out that the performance of the variance ratio
test is severely affected by the aggregated result of non-trading
and portfolio effects. However, it may be interesting to analyze how
the above-mentioned effects alter the statistical distribution of
the parameters associated with the asset return time series.\\

In order to appreciate the possible effect of non-trading on certain
parameters such as (a) asset correlation, (b) asset autocorrelation,
and (c) asset cross-autocorrelation, a simulation of 100.000 trajectories
based on a normal multivariate return series (of 5.000 observations
each) was carried out. Simulations were initially generated (without
non-trading effects) with a returns correlation between assets of
90$\%$.\\

Figure 21 shows that the histogram of the correlation observed without
non-trading effect (blue) has a mean of 90$\%$ and a standard deviation
of 0.006. The histogram also shows the correlation between the assets
when non-trading effects are introduced to one of the returns (red),
with a non-trading probability equal to 20$\%$. In this last case,
the average correlation was 0.689, and the standard deviation was
0.0271.

\begin{figure}[H]
\centering \includegraphics[width=0.7\linewidth]{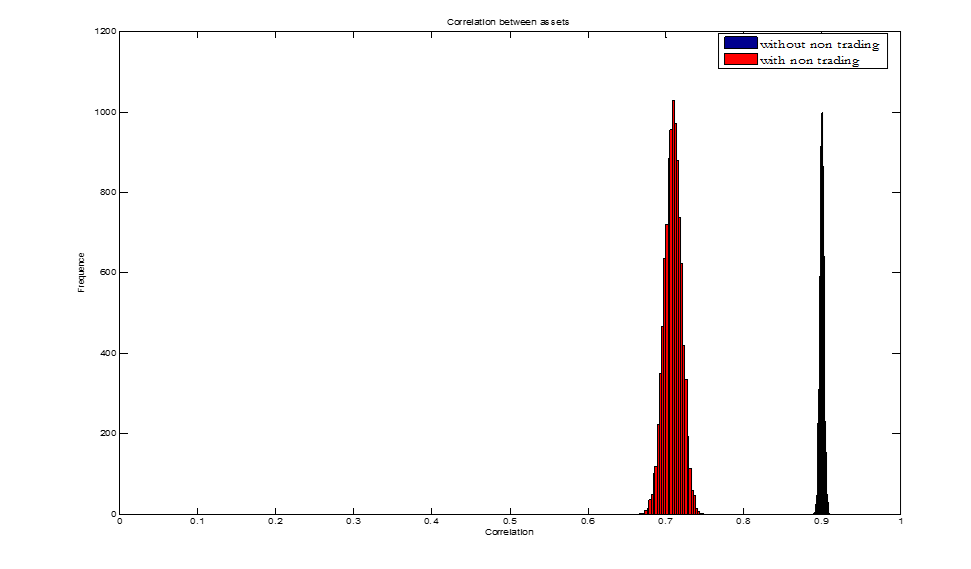} \protect\caption{Correlation between assets with and without nontrading}

\label{fig:Fig21} 
\end{figure}

\begin{figure}[H]
\centering \includegraphics[width=0.7\linewidth]{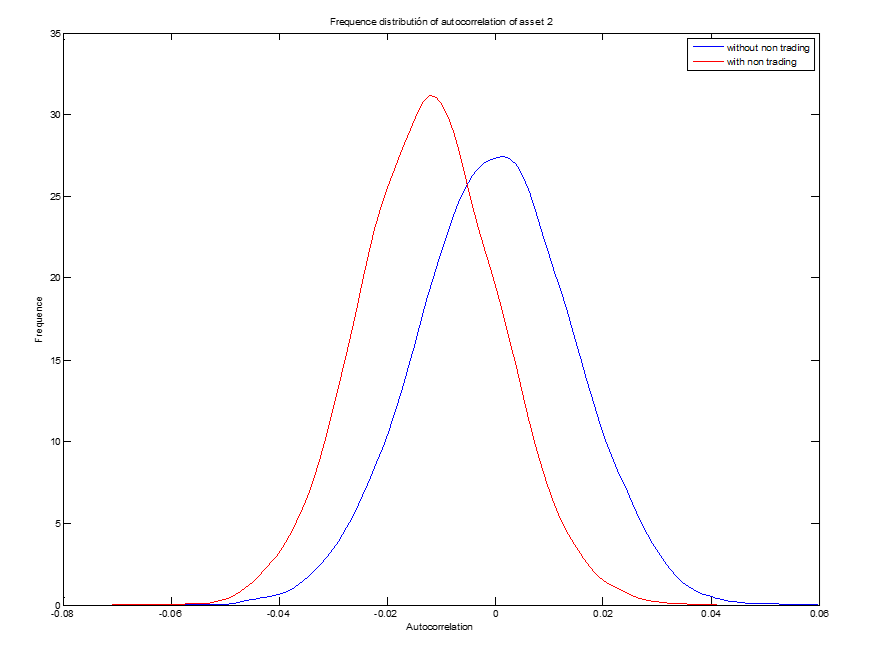} \protect\caption{Frequency distribution of autocorrelation of asset 2}

\label{fig:Fig22} 
\end{figure}

Figure 22 shows that autocorrelation observed in an asset with non-trading
was slightly negative. The average autocorrelation without nontrading
was -0.001, and the one with non-trading was -0.0453.

\begin{figure}[H]
\centering \includegraphics[width=0.7\linewidth]{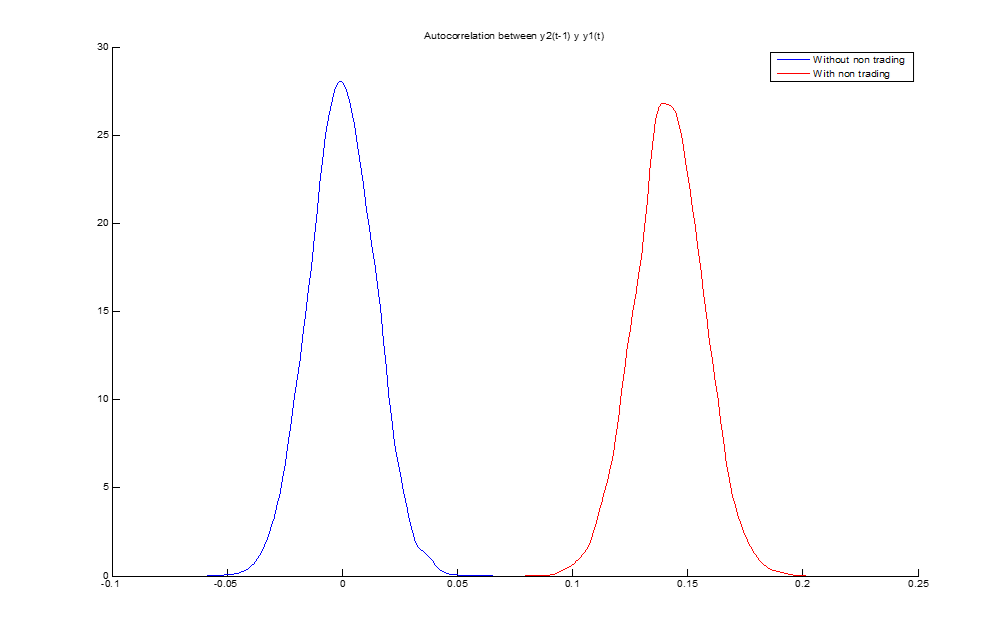} \protect\caption{Frequency distribution of cross-autocorrelation between assets}

\label{fig:Fig23} 
\end{figure}

Figure 23 shows the cross-autocorrelation between asset returns $r_{2}^{o}(t)$
and $r_{1}^{o}(t-1)$. The blue and red curves show the frequency
distribution of cross-autocorrelation without and with non-trading,
respectively. Mean crossed autocorrelation without non-trading is
approximately zero, and it has a variation range between -0.05 and
0.05. Mean with nontrading is 14.2$\%$, with a variation range between
0.09 and 0.2.

\subsection{Making sense of the experiment results}

As mentioned previously, to prove the random walk hypothesis (for
the stock market index) using a variance ratio test is equivalent
to showing that portfolio autocorrelations are zero. As was observed,
the rejection of the null hypothesis in the index was due to positive
asset autocorrelation. To illustrate how non-synchronous trading can
induce positive autocorrelation in a portfolio, consider the following
expression for a portfolio autocorrelation:

\begin{equation}
\rho=\rho(r_{p}(t),r_{p}(t-1))=\frac{cov(r_{p}(t),r_{p}(t-1))}{var(r_{p}(t))}
\end{equation}
where \ $r_{p}=\sum_{i=1}^{n}\alpha_{i}r_{i}(t)$ \ subject to \
$\sum_{i=1}^{n}\alpha_{i}=1$, that is {\small{}
\begin{equation}
{\rho=\frac{\sum_{i=1}^{n}\alpha_{i}^{2}Cov(r_{i(t)},r_{i(t-1))}+\sum_{=i=1}^{n}{\substack{\sum_{j=1}^{n}\\
j\neq i
}
\alpha_{i}\alpha_{j}Cov(r_{i(t)},r_{j(t))})}}{\sum_{i=1}^{n}V(r_{i}(t))+\sum_{=i=1}^{n}{\substack{\sum_{j=1}^{n}\\
j\neq i
}
\alpha_{i}\alpha_{j}Cov(r_{i(t)},r_{j(t)})}}}
\end{equation}
} Considering the case of an index that contains only two financial
assets, the autocorrelation of portfolio is: {\tiny{}
\begin{equation}
{\rho=\frac{\alpha^{2}Cov(r_{1(t)},r_{1(t-1)})+(1-\alpha)^{2}Cov(r_{2(t)},r_{2(t-1)})+\alpha(1-\alpha)Cov(r_{1(t)},r_{2(t-1)})+\alpha(1-\alpha)Cov(r_{2(t)},r_{1(t-1))})}{\alpha^{2}V(r_{1(t)})+(1-\alpha)^{2}V(r_{2(t)})+2\alpha(1-\alpha)Cov(r_{1(t)}),r_{2(t))}}}
\end{equation}
} To see the incidence of non-synchronous trading in this portfolio,
we introduced the non-trading effect on the second asset and then
observed what happened with estimated values for some of the most
important parameters of the portfolio returns. We assumed that the
virtual returns (without non-synchronous trading effect) had a multivariate
normal distribution with parameter $\mu=[0.005,0.01]$ and 

\begin{center}
$V=\begin{bmatrix}0.005^{2} & 0.09x0.005x0.02\\
0.09x0.005x0.02 & 0.02^{2}
\end{bmatrix}$ 
\par\end{center}

Using these stochastic processes, a simulation of 10.000 trajectories
was generated (each one of 5.000 periods) without non-trading. Mean
values were calculated for cross-autocorrelations and correlations
between individual assets. Also, standard deviations were computed.
On average, the autocorrelations were equal to zero, and correlations
between assets and variance between assets remained the same. The
average values of these parameters is shown in row 2 of Table 5. Next,
we will perform the same exercise, but this time we will introduce
a 20$\%$ non-trading to Asset 2. The average values of this parameter
is shown in row 3 of Table 5.

{\tiny{}}
\begin{table}[H]
{\tiny{}}%
\begin{tabular}{|c|lllllll|}
\hline 
{\tiny{}nontrading } & {\tiny{}$\sigma_{r_{1}}$} & {\tiny{}$\sigma_{r_{2}}$} & {\tiny{}$\rho_{r_{1(t)},r_{2(t)}}$} & {\tiny{}$\rho_{r_{1(t)},r_{2(t-1)}}$ } & {\tiny{}$\rho_{r_{2(t)},_{1(t-1)}}$ } & {\tiny{}$\rho_{r_{1(t)},r_{1(t-1)}}$} & {\tiny{}$\rho_{r_{2(t)},r_{2(t-1)}}$}\tabularnewline
\hline 
{\tiny{}Without} & {\tiny{}0.05 } & {\tiny{}0.02 } & {\tiny{}0.9 } & {\tiny{}0.0 } & {\tiny{}-0.0009 } & {\tiny{}0 } & {\tiny{}-0.0011}\tabularnewline
\hline 
{\tiny{}With } & {\tiny{}0.05 } & {\tiny{}0.02 } & {\tiny{}0.689 } & {\tiny{}-0.0008 } & {\tiny{}0.1349 } & {\tiny{}0 } & {\tiny{}-0.0453}\tabularnewline
\hline 
\end{tabular}{\tiny{}\protect\caption{{\tiny{}Correlations and variance between assets}}
}{\tiny \par}

{\tiny{}\label{Tabla 5} }
\end{table}
{\tiny \par}

 As shown in Table 5, non-trading on the second asset generates: 
\begin{description}
\item [{a)}] A slight increase in the standard deviation of asset 2; 
\item [{b)}] A reduction in the asset correlation (from 0.8999 to 0.689); 
\item [{c)}] A significant rise in cross-autocorrelation between assets
2 and 1, growing from -0.0009 to 0.1349; and 
\item [{d)}] A negative autocorrelation in asset 2. 
\end{description}
\noindent The previous example demonstrated that non-synchronous trading
can induce negative autocorrelation in individual assets and it can
cause positive cross-autocorrelation between the assets. Furthermore,
it can induce positive autocorrelation in the stock market index.
With the calculated parameter one can compute the portfolio autocorrelation
using equation (25). These values are shown in Figure 24.

\begin{figure}[H]
\centering \includegraphics[width=0.7\linewidth]{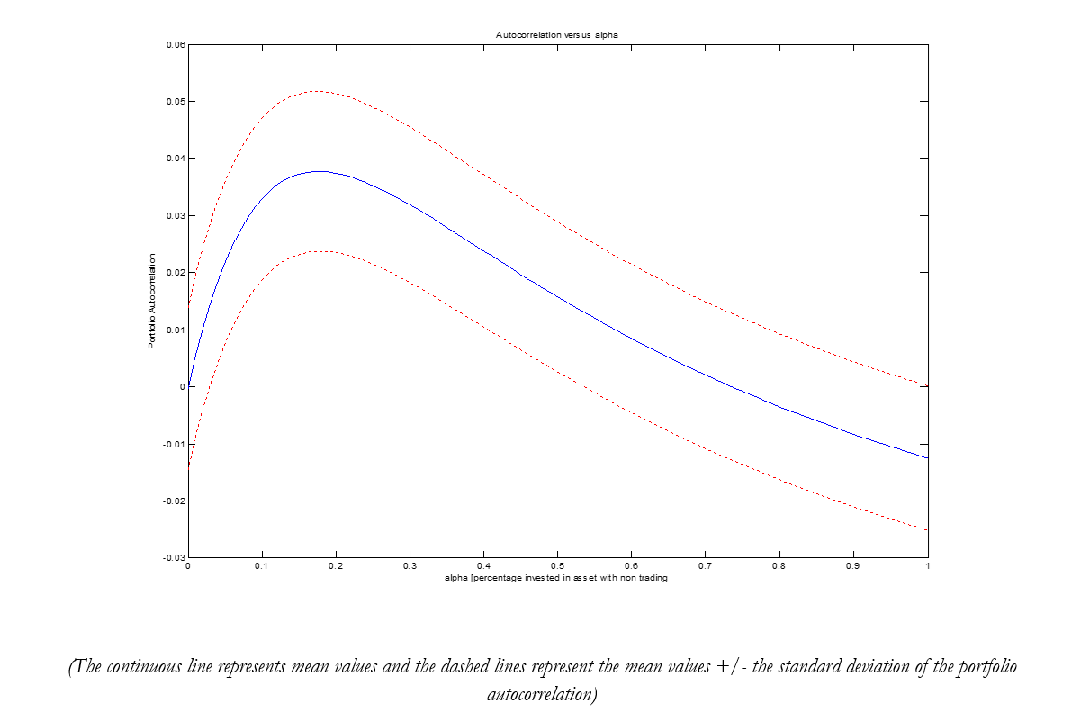} \protect\caption{Autocorrelation of portfolio versus percentage invested in asset with
non-trading}

\label{fig:Fig24} 
\end{figure}

Figure 24 shows that the autocorrelation induced by non-synchronous
trading is considerably higher than zero for some portfolios. The
blue line plots the average portfolio autocorrelation, and the red
dotted lines plot the mean of the portfolio autocorrelation +/- twice
its standard deviation. The intensity of this phenomenon increases
when the asset number increases within the portfolio.

\subsection{Weekly portfolio returns autocorrelation estimation. Six-asset cases. }

Hereafter, the estimation of theoretical autocorrelation induced by
non-synchronous trading was undertaken considering a six-asset portfolio
that had:
\begin{description}
\item [{a)}] Heterogeneous features, and 
\item [{b)}] Non-trading daily probabilities equal to those used by \citet{boudoukh1994tale}.
\end{description}
In this way, non-trading time series of daily returns were generated.
Then, weekly returns were calculated as the sum of 5 daily returns
and from them weekly autocorrelations of different portfolios were
estimated. The portfolios were built explicitly as%
\footnote{There are multiple ways to construct portfolios with six assets. However,
in this case, to illustrate the non-trading and portfolio effects,
we chose this particular method.%
} 
\begin{equation}
r_{p}^{o}=(1-\alpha)\frac{r_{1}^{o}+r_{2}^{o}+r_{3}^{o}}{3}+\alpha\frac{r_{4}^{o}+r_{5}^{o}+r_{6}^{o}}{3}
\end{equation}
Table 6 shows the values of parameters used. We assumed that assets
1, 2, and 3 had characteristics similar to assets of bigger firms
and that Assets 4, 5, and 6 had characteristics similar to assets
of smaller firms%
\footnote{$\pi_{i}$ is the non-trading probability%
}.

\begin{table}
\centering \protect\caption{Values of the parameters used in the simulation}

\label{my-label} %
\begin{tabular}{lrrrrrrrrr}
\multicolumn{4}{l}{} & \multicolumn{6}{c}{Correlation Matrix}\tabularnewline
\multicolumn{4}{l}{} & \multicolumn{6}{c}{Assets}\tabularnewline
Asset  & \multicolumn{1}{l}{$\pi_{i}$} & \multicolumn{1}{l}{Return} & \multicolumn{1}{l}{Volatility} & \multicolumn{1}{l}{1} & \multicolumn{1}{l}{2} & \multicolumn{1}{l}{3} & \multicolumn{1}{l}{4} & \multicolumn{1}{l}{5} & \multicolumn{1}{l}{6}\tabularnewline
1  & 0\%  & 0.5\%  & 0.5\%  & 1  & 0.9  & 0.9  & 0.9  & 0.9  & 0.9 \tabularnewline
2  & 0\%  & 0.5\%  & 1\%  & 0.9  & 1  & 0.9  & 0.9  & 0.9  & 0.9 \tabularnewline
3  & 0\%  & 0.5\%  & 1.5\%  & 0.9  & 0.9  & 1  & 0.9  & 0.9  & 0.9 \tabularnewline
4  & 43\%  & 0.5\%  & 2\%  & 0.9  & 0.9  & 0.9  & 1  & 0.9  & 0.9 \tabularnewline
5  & 60\%  & 0.5\%  & 2.5\%  & 0.9  & 0.9  & 0.9  & 0.9  & 1  & 0.9 \tabularnewline
6  & 85\%  & 0.5\%  & 3\%  & 0.9  & 0.9  & 0.9  & 0.9  & 0.9  & 1 \tabularnewline
\end{tabular}
\end{table}

The simulation was done for a sample series of 10.000 daily returns,
which consisted of 5.000 time points. Figure 25 shows the weekly autocorrelation
induced by non-trading for the six-asset portfolio%
\footnote{The continuous blue line represents the mean values, and the dashed
red lines the mean values $+/-$ two standard deviations associated
with the portfolio autocorrelation%
}.

\begin{figure}[H]
\centering \includegraphics[width=1\linewidth]{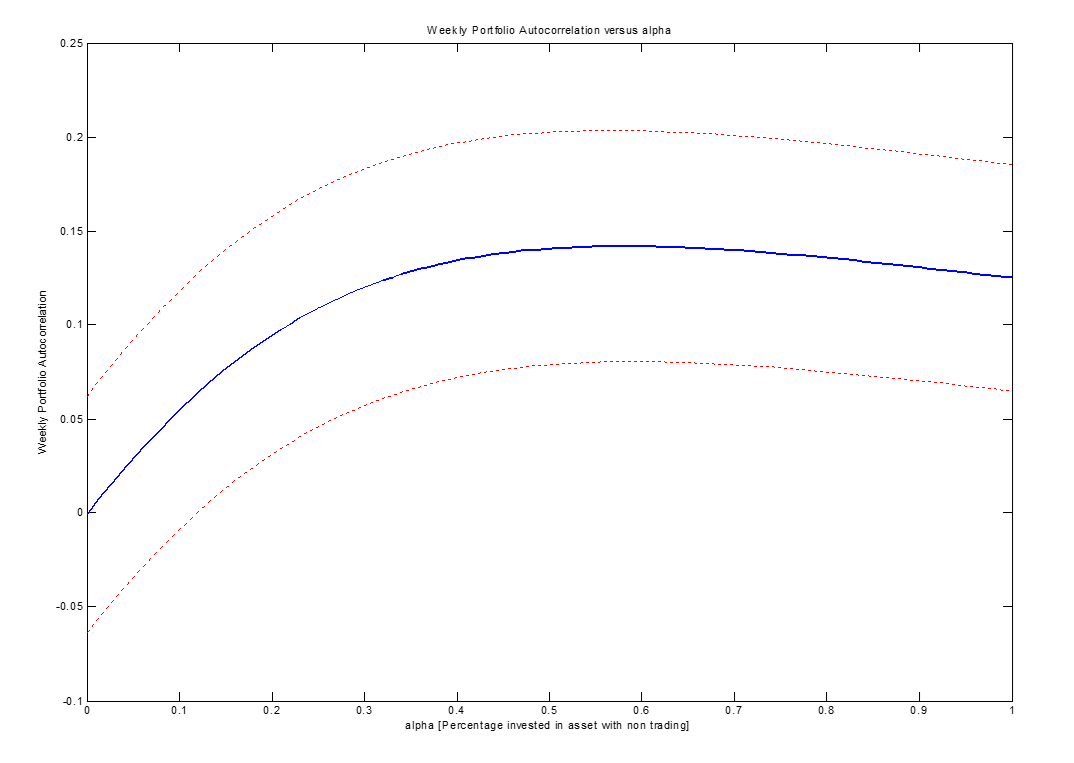} \protect\caption{Weekly autocorrelation of portfolio versus percentage invested in
asset with nontrading.}

\label{fig:Fig25} 
\end{figure}

As can be seen in Figure 25, the weekly portfolio autocorrelation
depended on the portfolio structure. However, it is possible to see
portfolios in which weekly autocorrelation induced by non-synchronous
trading can reach values of 20$\%$ for some portfolios and autocorrelations
lower than -5$\%$ for other ones. Given the latter facts, we can
point out that the joint effect of portfolio and non-synchronous trading
can explain both: 
\begin{description}
\item [{a)}] The high autocorrelations empirically observed in the CRSP
NYSE equal weighted index ($\alpha\approx0.5$), and 
\item [{b)}] The lower autocorrelations for the CRSP NYSE value-weighted
index ($\alpha\approx0$) 
\end{description}

\section{Conclusions}

In this paper, we analyzed the historical evolution of the acceptance
or rejection of the random walk hypothesis applied to the main stock
exchange index of NYSE. It has been shown that the random walk hypothesis
is rejected for each of the windows for 1601 weeks from 1950-2013
for the equally weighted index. However, for the index weighted by
market value, the hypothesis fails to be rejected from the late 1990s.
This shows a significant change in the behavior of the stock market
in New York. \\

The posibility that this divergent behavior between these index ought
to be explained by the combined effect of non-synchronous trading
and the portfolio, was explored. To investigate the feasibility of
this explanation we simulated the effects of these factors over two
and six assets portfolios . The samples were of finite size and the
characteristics of assets were assumed to be heterogeneous.\\
 It was found that the combined non-trading and portfolio effects
seriously affected the performance of the variance ratio test. For
the case of the two-asset portfolio, the only case in which the variance
ratio test outcomes were independent of the non-trading process was
when the correlation parameter was close to zero.\\

The results show that when the correlation between assets was not
zero, the value of the statistical variance ratio test depended explicitly
on the way the asset portfolio was built, that is, it presented an
$\alpha$ dependence. \\

The same conclusions were obtained for the six-asset portfolio. In
addition, it was found that the heterogeneity in the characteristics
of the individual assets increased the autocorrelation induced by
the non-synchronous trading.\\

Lastly, we can point out that the joint effect of portfolio and non-synchronous
trading can explain both:
\begin{description}
\item [{a)}] The high autocorrelations empirically observed in the CRSP
NYSE equal weighted index ($\alpha\approx0.5$), and 
\item [{b)}] The lower autocorrelations for the CRSP NYSE value-weighted
index ($\alpha\approx0$) 
\end{description}
\pagebreak{}\bibliographystyle{plainnat}
\bibliography{BiblioPaperA}

\begin{thebibliography}{37}
\providecommand{\natexlab}[1]{#1}
\providecommand{\url}[1]{\texttt{#1}}
\expandafter\ifx\csname urlstyle\endcsname\relax
  \providecommand{\doi}[1]{doi: #1}\else
  \providecommand{\doi}{doi: \begingroup \urlstyle{rm}\Url}\fi

\bibitem[Atchison et~al.(1987)Atchison, Butler, and
  Simonds]{atchison1987nonsynchronous}
Michael~D Atchison, Kirt~C Butler, and Richard~R Simonds.
\newblock Nonsynchronous security trading and market index autocorrelation.
\newblock \emph{Journal of Finance}, pages 111--118, 1987.

\bibitem[Bachelier(1900)]{bachelier1900theorie}
Louis Bachelier.
\newblock \emph{Th{\'e}orie de la sp{\'e}culation}.
\newblock Gauthier-Villars, 1900.

\bibitem[Bollerslev(1986)]{bollerslev1986generalized}
Tim Bollerslev.
\newblock Generalized autoregressive conditional heteroskedasticity.
\newblock \emph{Journal of econometrics}, 31\penalty0 (3):\penalty0 307--327,
  1986.

\bibitem[Boudoukh et~al.(1994)Boudoukh, Richardson, and
  Whitelaw]{boudoukh1994tale}
Jacob Boudoukh, Matthew~P Richardson, and RE~Whitelaw.
\newblock A tale of three schools: Insights on autocorrelations of
  short-horizon stock returns.
\newblock \emph{Review of financial studies}, 7\penalty0 (3):\penalty0
  539--573, 1994.

\bibitem[Campbell et~al.(1997)Campbell, Lo, MacKinlay,
  et~al.]{campbell1997econometrics}
John~Y Campbell, Andrew Wen-Chuan Lo, Archie~Craig MacKinlay, et~al.
\newblock \emph{The econometrics of financial markets}, volume~2.
\newblock princeton University press Princeton, NJ, 1997.

\bibitem[Cecchetti and Lam(1994)]{cecchetti1994variance}
Stephen~G Cecchetti and Pok-sang Lam.
\newblock Variance-ratio tests: small-sample properties with an application to
  international output data.
\newblock \emph{Journal of Business \& Economic Statistics}, 12\penalty0
  (2):\penalty0 177--186, 1994.

\bibitem[Chelley-Steeley and Steeley(2014)]{chelley2014portfolio}
Patricia~L Chelley-Steeley and James~M Steeley.
\newblock Portfolio size, non-trading frequency and portfolio return
  autocorrelation.
\newblock \emph{Journal of international financial markets, institutions and
  money}, 33:\penalty0 56--77, 2014.

\bibitem[Chen and Deo(2006)]{chen2006variance}
Willa~W Chen and Rohit~S Deo.
\newblock The variance ratio statistic at large horizons.
\newblock \emph{Econometric Theory}, 22\penalty0 (02):\penalty0 206--234, 2006.

\bibitem[Chow and Denning(1993)]{chow1993simple}
K~Victor Chow and Karen~C Denning.
\newblock A simple multiple variance ratio test.
\newblock \emph{Journal of Econometrics}, 58\penalty0 (3):\penalty0 385--401,
  1993.

\bibitem[Cohen et~al.(1983)Cohen, Hawawini, Maier, Schwartz, and
  Whitcomb]{cohen1983friction}
Kalman~J Cohen, Gabriel~A Hawawini, Steven~F Maier, Robert~A Schwartz, and
  David~K Whitcomb.
\newblock Friction in the trading process and the estimation of systematic
  risk.
\newblock \emph{Journal of Financial Economics}, 12\penalty0 (2):\penalty0
  263--278, 1983.

\bibitem[Conrad and Kaul(1988)]{conrad1988time}
Jennifer Conrad and Gautam Kaul.
\newblock Time-variation in expected returns.
\newblock \emph{Journal of business}, pages 409--425, 1988.

\bibitem[Dimson(1979)]{dimson1979risk}
Elroy Dimson.
\newblock Risk measurement when shares are subject to infrequent trading.
\newblock \emph{Journal of Financial Economics}, 7\penalty0 (2):\penalty0
  197--226, 1979.

\bibitem[Efron(1979)]{efron1979bootstrap}
Bradley Efron.
\newblock Bootstrap methods: another look at the jackknife.
\newblock \emph{The annals of Statistics}, pages 1--26, 1979.

\bibitem[Engle(1982)]{engle1982autoregressive}
Robert~F Engle.
\newblock Autoregressive conditional heteroscedasticity with estimates of the
  variance of united kingdom inflation.
\newblock \emph{Econometrica: Journal of the Econometric Society}, pages
  987--1007, 1982.

\bibitem[Epps(1979)]{epps1979comovements}
Thomas~W Epps.
\newblock Comovements in stock prices in the very short run.
\newblock \emph{Journal of the American Statistical Association}, 74\penalty0
  (366a):\penalty0 291--298, 1979.

\bibitem[Fama(1965)]{fama1965behavior}
Eugene~F Fama.
\newblock The behavior of stock-market prices.
\newblock \emph{Journal of business}, pages 34--105, 1965.

\bibitem[Fama(1970)]{fama1970efficient}
Eugene~F Fama.
\newblock Efficient capital markets: A review of theory and empirical work.
\newblock \emph{The journal of Finance}, 25\penalty0 (2):\penalty0 383--417,
  1970.

\bibitem[Fama(1991)]{fama1991efficient}
Eugene~F Fama.
\newblock Efficient capital markets: Ii.
\newblock \emph{The journal of finance}, 46\penalty0 (5):\penalty0 1575--1617,
  1991.

\bibitem[Fisher(1966)]{fisher1966some}
Lawrence Fisher.
\newblock Some new stock-market indexes.
\newblock \emph{Journal of Business}, pages 191--225, 1966.

\bibitem[Grossman and Stiglitz(1980)]{grossman1980impossibility}
Sanford~J Grossman and Joseph~E Stiglitz.
\newblock On the impossibility of informationally efficient markets.
\newblock \emph{The American economic review}, pages 393--408, 1980.

\bibitem[Jensen(1978)]{jensen1978some}
Michael~C Jensen.
\newblock Some anomalous evidence regarding market efficiency.
\newblock \emph{Journal of financial economics}, 6\penalty0 (2):\penalty0
  95--101, 1978.

\bibitem[Kan(2006)]{kan2006exact}
Raymond Kan.
\newblock Exact variance ratio test with overlapping data.
\newblock \emph{Available at SSRN 891680}, 2006.

\bibitem[Kanellopoulou and Panas(2008)]{kanellopoulou2008empirical}
Stella Kanellopoulou and Epaminondas Panas.
\newblock Empirical distributions of stock returns: Paris stock market,
  1980--2003.
\newblock \emph{Applied Financial Economics}, 18\penalty0 (16):\penalty0
  1289--1302, 2008.

\bibitem[Kim(2006)]{kim2006wild}
Jae~H Kim.
\newblock Wild bootstrapping variance ratio tests.
\newblock \emph{Economics letters}, 92\penalty0 (1):\penalty0 38--43, 2006.

\bibitem[Lo and MacKinlay(1988)]{lo1988stock}
Andrew~W Lo and A~Craig MacKinlay.
\newblock Stock market prices do not follow random walks: Evidence from a
  simple specification test.
\newblock \emph{Review of financial studies}, 1\penalty0 (1):\penalty0 41--66,
  1988.

\bibitem[Lo and MacKinlay(1990)]{lo1990econometric}
Andrew~W Lo and A~Craig MacKinlay.
\newblock An econometric analysis of nonsynchronous trading.
\newblock \emph{Journal of Econometrics}, 45\penalty0 (1):\penalty0 181--211,
  1990.

\bibitem[Mandelbrot(1997)]{mandelbrot1997variation}
Benoit~B Mandelbrot.
\newblock \emph{The variation of certain speculative prices}.
\newblock Springer, 1997.

\bibitem[Mech(1993)]{mech1993portfolio}
Timothy~S Mech.
\newblock Portfolio return autocorrelation.
\newblock \emph{Journal of Financial Economics}, 34\penalty0 (3):\penalty0
  307--344, 1993.

\bibitem[Osborne(1959)]{osborne1959brownian}
MF~Maury Osborne.
\newblock Brownian motion in the stock market.
\newblock \emph{Operations research}, 7\penalty0 (2):\penalty0 145--173, 1959.

\bibitem[Perry(1985)]{perry1985portfolio}
Philip~R Perry.
\newblock Portfolio serial correlation and nonsynchronous trading.
\newblock \emph{Journal of Financial and Quantitative Analysis}, 20\penalty0
  (04):\penalty0 517--523, 1985.

\bibitem[Richardson and Smith(1991)]{richardson1991tests}
Matthew Richardson and Tom Smith.
\newblock Tests of financial models in the presence of overlapping
  observations.
\newblock \emph{Review of Financial Studies}, 4\penalty0 (2):\penalty0
  227--254, 1991.

\bibitem[Robert(1967)]{Robert1967}
Robert.
\newblock Statistical versus clinical prediction of the stock market.
\newblock \emph{Unpublished manuscript}, 1967.

\bibitem[Scholes and Williams(1977)]{scholes1977estimating}
Myron Scholes and Joseph Williams.
\newblock Estimating betas from nonsynchronous data.
\newblock \emph{Journal of financial economics}, 5\penalty0 (3):\penalty0
  309--327, 1977.

\bibitem[Whang and Kim(2003)]{whang2003multiple}
Yoon-Jae Whang and Jinho Kim.
\newblock A multiple variance ratio test using subsampling.
\newblock \emph{Economics Letters}, 79\penalty0 (2):\penalty0 225--230, 2003.

\bibitem[White(1980)]{white1980heteroskedasticity}
Halbert White.
\newblock A heteroskedasticity-consistent covariance matrix estimator and a
  direct test for heteroskedasticity.
\newblock \emph{Econometrica: Journal of the Econometric Society}, pages
  817--838, 1980.

\bibitem[White and Domowitz(1984)]{white1984nonlinear}
Halbert White and Ian Domowitz.
\newblock Nonlinear regression with dependent observations.
\newblock \emph{Econometrica: Journal of the Econometric Society}, pages
  143--161, 1984.

\bibitem[Wright(2000)]{wright2000alternative}
Jonathan~H Wright.
\newblock Alternative variance-ratio tests using ranks and signs.
\newblock \emph{Journal of Business \& Economic Statistics}, 18\penalty0
  (1):\penalty0 1--9, 2000.

\end{thebibliography}
 \bibliographystyle{apalike}
\end{document}